\documentclass[lineno]{jfm}

\makeatletter
\gdef\@underjournal{\vbox to 5.5\p@{\vss}}
\makeatother

\usepackage{graphicx}
\usepackage{newtxtext}
\usepackage{newtxmath}
\usepackage{natbib}
\usepackage{tikz}
\usepackage{soul}
\usetikzlibrary{arrows,decorations.markings}
\usetikzlibrary{arrows.meta}
\usetikzlibrary{positioning,shapes}
\usepackage{hyperref}
\hypersetup{
    colorlinks = true,
    urlcolor   = blue,
    citecolor  = black,
}
\usepackage{calc}

\newcommand{\RomanNumeralCaps}[1]

\usepackage{epstopdf,epsfig}
\usepackage{amsmath}
\usepackage{psfrag}
\usepackage{subfigure}
\usepackage{mathrsfs}
\usepackage{color}
\usepackage{overpic}
\usepackage{soul}
\usepackage{tikz}
\usepackage{comment}
\usepackage[dvipsnames]{xcolor}
\definecolor{DarkGreen}{RGB}{0, 128, 0}

\shorttitle{Instability of cooled high-supersonic streaks}
\shortauthor{D. Xu, P. Ricco, E. Marensi and X. Wu}

\title{Instability of streaky high-supersonic boundary layers over cooled walls}
\author{Dongdong Xu\aff{1}\corresp{\email{dongdongxu01@outlook.com}}, 
Pierre Ricco\aff{1},
Elena Marensi\aff{1} 
and 
Xuesong Wu\aff{2}
}

\affiliation{
\aff{1}
School of Mechanical, Aerospace and Civil Engineering, The University of Sheffield, Sheffield, S1 3JD, United Kingdom
\aff{2}
Department of Mathematics, Imperial College London, London, SW7 2AZ, United Kingdom
}

\begin{document}
\nolinenumbers
\let\runninglinenumbers\relax
\maketitle

\begin{abstract}
We investigate the instability of streaks induced by free-stream vortical disturbances in cooled-wall high-supersonic boundary layers. Distinct types of instability modes are identified: modified Mack second modes, a newly discovered varicose secondary-instability mode and sinuous or varicose secondary-instability modes akin to those in incompressible boundary layers. Unique characteristics of the new secondary-instability mode are that its eigenfunctions concentrate in valley regions and feature intense temperature fluctuations near the wall. The streaks in high-supersonic boundary layers destabilise the three-dimensional secondary-instability modes, but suppress the growth of the modified Mack second modes, both radiating and non-radiating. The suppression of radiating modes amounts to a reduction of the two-dimensional noise they emit. Numerical results indicate that, unlike streaks in adiabatic-wall boundary layers, whose instability concentrates in low-speed peak regions away from the wall, the secondary instability of streaks in cooled-wall high-supersonic boundary layers occurs both in low-speed peak regions and in high-speed valley regions.
\end{abstract}

\vspace{-15mm}
\section{Introduction}
Transition to turbulence in supersonic boundary layers poses significant challenges in the design of high-speed vehicles because it causes an undesired increase in wall friction and wall-heat transfer, compromising vehicle performance \citep*{schneider1999flight}. It is therefore imperative to understand how and where supersonic boundary-layer transition occurs and to find effective means of delaying the breakdown to turbulence.

Supersonic transition is greatly influenced by free-stream disturbances, which are inevitably present in laboratory and engineering systems. These disturbances consist of vortical, acoustic and entropy modes, of which free-stream vortical disturbances (FVD) are most likely to excite streaks \citep*{leib1999effect}. As in incompressible boundary layers, transition to turbulence in supersonic boundary layers takes place through different mechanisms depending on the intensity of FVD. For low-intensity FVD, transition first starts with the amplification of two-dimensional Mack second modes, which undergo a fundamental type of parametric secondary instability, leading to the formation of streaks \citep{hader2020three}. For sufficiently intense FVD,  streaks are generated directly near the leading edge of the flat plate and amplify under the continued influence of FVD. The transition scenario that ensues from the secondary instability of these streaks is termed bypass transition.
A rigorous mathematical framework for the excitation of streaks by weak FVD in the incompressible regime was formulated by \cite{leib1999effect}. They showed that the evolution of streaks is governed by the boundary-region equations, i.e. the asymptotic limit of the Navier-Stokes equations for low-frequency disturbances. In the boundary-region framework, the interaction between FVD and a boundary layer is accounted for through appropriate initial and free-stream boundary conditions. \cite{ricco2011evolution} and \cite{marensi2017nonlinear} extended the approach of \cite{leib1999effect} to investigate the generation of nonlinear streaks by FVD in incompressible and compressible boundary layers, respectively. \cite{paredes2019instability} chose instead to model streaks in supersonic and hypersonic boundary layers as disturbances that develop from linear optimal perturbations \citep{andersson2001breakdown}. Following a similar approach, \cite{caillaud2025} analysed the linear instability of streaks generated by optimal disturbances in a hypersonic boundary layer.

Wall cooling of supersonic and hypersonic boundary layers is often used in flight conditions to delay transition and provide thermal protection \citep{xu2023review}. The wall temperature \(T_w\) is typically much lower than the adiabatic wall temperature \(T_{ad}\) ($T_w/T_{ad} < 0.3$, \citeauthor{choudhari2009transition}~\citeyear{choudhari2009transition}). The impact of wall cooling on boundary-layer transition is multifaceted, influencing the instability of two-dimensional modes, the generation of three-dimensional streaks through receptivity to FVD, and the related secondary instability \citep{surujhlal2024hot}. Wall cooling destabilises the two-dimensional Mack second modes \citep{mack1984boundary,little2025} and reduces their phase speeds \citep{fedorov2003initial}. Furthermore, cooled walls induce radiating modes that propagate supersonically relative to the free stream, i.e. sound waves are emitted into the far field \citep{mack1984boundary,bitter2015stability,knisely2019sounda}. These findings were confirmed by direct numerical simulations (DNS) \citep{surujhlal2024hot,chuvakhov2016spontaneous,knisely2019soundb}. DNS results also indicated that beyond the linear regime, the intensified mean-flow gradients and baroclinicity due to wall cooling persist into the nonlinear stages, affecting the evolution of instability wave packets \citep{salemi2018synchronization}. \cite{knisely2019soundb} further investigated intermodal dynamics. They observed that, for the case of moderate cooling, linearly damped supersonic modes amplify weakly through nonlinear interactions with coexisting waves in the continuous slow acoustic spectrum. Several experimental studies have investigated supersonic and hypersonic boundary-layer instability mechanisms in high-enthalpy facilities with various degrees of wall cooling \citep{schneider2008development,xu2023review}. The effect of streamwise localised wall cooling was also studied experimentally in hypersonic boundary layers and was found to reduce the amplitude of the second mode and delay transition \citep{fedorov2015high,sidorenko2015effect}.

Although the impact of wall cooling on the Mack-mode instability has received widespread attention, studies on supersonic boundary layers populated by streaky structures remain scarce. \cite{marensi2017nonlinear} examined streaks induced by FVD in a Mach 3 boundary layer and \cite{liu2022effects} investigated the secondary instability of these streaks under adiabatic conditions. The dynamics of the Mack second modes in streaky boundary layers was not considered in those studies. \cite{paredes2019instability} investigated the stabilising effects of streaks on the second-mode instability in the presence of wall cooling with the streaks being modelled by the so-called optimal perturbations. The information provided in these studies is inadequate for two principal reasons. First, the adiabatic condition utilised in \cite{liu2022effects} does not fully represent the thermal protection technology that must be deployed to reduce the surface temperature of hypersonic vehicles to a level sufficiently lower than the adiabatic value. Second, in flight and laboratory conditions, streaks are induced by naturally present external disturbances, which are not taken into account in the optimal-disturbance approach employed in \cite{paredes2019instability}. Optimal disturbances are computed by solving an optimisation procedure that gives rise to the maximum energy growth among all possible initial perturbations with equal amplitude. The optimal disturbances so identified represent the `worst scenario', but it is still debated whether and how the required wall-normal distributions at a given streamwise location could be created in a laboratory experiment. Nevertheless, optimal streaks can capture some qualitative features of those disturbances induced by naturally present external perturbations. They were thus used as a proxy for FVD-induced streaks in order to compute possible secondary-instability modes. However, quantitative relations between the characteristics of external disturbances, the associated stability properties and the transition location  cannot be established because FVD are absent in the optimal-disturbance approach. Evidence of these limitations was provided by the streaks induced by three-dimensional roughness arrays exhibiting wall-normal distributions and streamwise evolutions that were different from the optimal streaks \citep{white2002transient}. The roughness-induced streaks were also found to be more prone to secondary instability than the optimal streaks \citep{denissen2013secondary}. This result suggests that, despite leading to maximum amplification, the optimal disturbances may not represent the worst scenario of earliest transition.

In view of the discussion above, our central aim is to explain the influence of FVD-induced streaks on the Mack second mode and the ensuing secondary instability under the effect of wall cooling. Following \cite{marensi2017nonlinear}, a combination of asymptotic and numerical methods is employed to describe the excitation and nonlinear development of streaks in high-supersonic boundary layers over cooled walls. Unlike \cite{marensi2017nonlinear}, the present study focuses on the high-supersonic regime and examines the effects of wall cooling on both the streaky base flow and its secondary-instability modes. A simplified model of FVD is used to investigate the fundamental mechanisms rather than to reproduce a specific experiment quantitatively. We first examine the effect of wall cooling and FVD amplitude on the nonlinear supersonic streaky structures and show how they alter the growth rate of the Mack second modes, which are present even in the absence of streaks. The second objective is to study how the FVD amplitude and wall cooling impact the three-dimensional secondary instability, which arises only when FVD are sufficiently intense. We discover a new varicose mode that attains growth rates comparable to those of other modes and is destabilised by wall cooling. Within our unified receptivity--instability framework \citep{marensi2017nonlinear,xu2024excitation}, we assess the relative significance of the modified Mack second mode and the secondary-instability modes with respect to the FVD intensity.

In \S\ref{sec:methodology}, we describe the boundary-region framework governing the excitation and nonlinear evolution of the streaks, formulate the linear instability problem of the streaky boundary layer and present the numerical procedures for solving the boundary-region equations and the eigenvalue system. The results on the dynamics of the streaky structures and their stability are presented in \S\ref{sec:results}. The conclusions are given in \S\ref{sec:conclusions}.

\vspace{-4mm}
\section{Methodology}
\label{sec:methodology}
We consider a supersonic boundary layer over a flat plate. The oncoming base flow is uniform with velocity $U_\infty^*$ and temperature $T_\infty^*.$ The superscript $*$ denotes dimensional quantities, while quantities not marked by any symbol are non-dimensional. The flow is described by the Cartesian coordinate system $\boldsymbol{x}^*=(x^*,y^*,z^*)$ defining the streamwise, wall-normal and spanwise directions. Although free-stream turbulence is of broadband nature, we consider the simplified case of FVD consisting of a pair of vortical modes with the same frequency (and hence streamwise wavenumber), but opposite spanwise wavenumbers $\pm k_z^*$, as in \cite{ricco2011evolution} and \cite{marensi2017nonlinear}. Velocities are scaled by $U_\infty^*$ and temperatures are scaled by $T_\infty^*.$ Lengths are normalised by $\Lambda^*=\lambda_z^*/2\pi$, where $\lambda_z^*$ is the spanwise wavelength of the FVD. The FVD are expressed as 
\[\boldsymbol{u}-{\rm \bf{i}}= \epsilon \left(\boldsymbol{\hat u}_+^\infty {\rm e}^{i k_z z} +\boldsymbol{\hat u}_-^\infty {\rm e}^{-i k_z z}\right) {\rm e}^{i k_x (x-t) + i k_y y}+{\rm c.c.},\]
where $\epsilon\ll 1$ is a measure of the FVD intensity, ${\rm \bf{i}}$ is the unit vector along $x$ and $\rm c.c.$ indicates the complex conjugate. 
The wavenumber vector $\boldsymbol{k}=\{k_x,k_y,k_z\}$ and the amplitude $\boldsymbol{\hat u}_{\pm}^\infty=\left\{{\hat u}_{x, \pm}^\infty,{\hat u}_{y, \pm}^\infty,{\hat u}_{z, \pm}^\infty\right\}=\mathcal{O}(1)$ satisfy the solenoidal condition 
\[k_x \hat{u}_{x,\pm}^{\infty} + k_y \hat{u}_{y,\pm}^{\infty} \pm k_z \hat{u}_{z,\pm}^{\infty}=0,
\]
which follows from the continuity equation.
The Reynolds number is $R_\Lambda={U_\infty^* \Lambda^*} /{\nu_\infty^*} \gg 1$ and the free-stream Mach number is $\mathcal{M}_\infty=U_\infty^*/a_\infty^*=\mathcal{O}(1),$ where $\nu^*_\infty$ and $a_\infty^*$ are the kinematic viscosity of air and the speed of sound in the free stream, respectively. The focus is on low-frequency disturbances because they penetrate the most into the boundary layer to generate streaks \citep{marensi2017nonlinear}. The slow temporal and spatial variables $\bar t= k_x t$ and $\bar x= k_x x$ are thus introduced. 

\subsection{Compressible boundary-region equations} 
The streamwise velocity component is larger than the wall-normal and spanwise components by a factor $\mathcal{O}(R_\Lambda)$ and exceeds the pressure perturbation by a factor $\mathcal{O}(R_\Lambda^2)$ \citep{marensi2017nonlinear}. The velocity, temperature and pressure are expressed as
\begin{eqnarray}
\dfrac{\left\{u^*,v^*,w^*\right\}}{U_\infty^*}=
\left\{\tilde u,\sqrt{\dfrac{k_x}{R_\Lambda}} \tilde v,k_x \tilde w\right\}, 
\;\;
\dfrac{T^*}{T_\infty^*}=\tilde T,
\;\;
\dfrac{p^*}{\rho^*_\infty U_\infty^{*2}}=\dfrac{k_x}{ R_\Lambda} \tilde p.
\label{eqn:u-p}
\end{eqnarray} 
Substituting expressions \eqref{eqn:u-p} into the compressible Navier-Stokes equations and performing the change of variables $(x, t) \to (\bar{x}, \bar{t})$ yields the leading-order nonlinear boundary-region equations:

\begin{equation}
\label{eqn:nubr-ta}
\dfrac{\p \tilde \rho}{\p \bar t}
+\dfrac{\p \tilde \rho \tilde u}{\p \bar x}
+\dfrac{1}{\chi}\dfrac{\p \tilde \rho \tilde v}{\p y}
+\dfrac{\p \tilde \rho \tilde w}{\p z}=0,
\end{equation}
\begin{equation}
\tilde \rho \dfrac{\p \tilde u}{\p \bar t}
+\tilde \rho \tilde u\dfrac{\p \tilde u}{\p \bar x}
+\tilde \rho \tilde v\dfrac{1}{\chi}\dfrac{\p \tilde u}{\p y}
+\tilde \rho \tilde w\dfrac{\p \tilde u}{\p z}=
\dfrac{1}{\chi^2}
\left[
\dfrac{\p }{\p y}
\left(
\tilde \mu\dfrac{\p \tilde u}{\p y}
\right)
+\dfrac{\p }{\p z}
\left(
\tilde \mu\dfrac{\p \tilde u}{\p z}
\right)
\right],
\end{equation}
\begin{equation}
\begin{split}
&
\tilde \rho\dfrac{\p \tilde v}{\p \bar t}
+\tilde \rho \tilde u\dfrac{\p \tilde v}{\p \bar x}
+\tilde \rho \tilde v\dfrac{1}{\chi}\dfrac{\p \tilde v}{\p y}
+\tilde \rho \tilde w\dfrac{\p \tilde v}{\p z}
=
\\ &
\dfrac{1}{\chi}
\left\{ -\dfrac{\p \tilde p}{\p y}
+
\dfrac{\p }{\p y}
\left[
\dfrac{2}{3}\tilde \mu
\left(
\dfrac{2}{\chi}\dfrac{\p \tilde v}{\p y}
-\dfrac{\p \tilde w}{\p z}
\right)
\right]
+
\dfrac{\p }{\p z}
\left[
\tilde \mu
\left(
\dfrac{1}{\chi}\dfrac{\p \tilde v}{\p z}
+\dfrac{\p \tilde w}{\p y}
\right)
\right]
-
\dfrac{\p }{\p y}
\left(
\dfrac{2}{3}\tilde \mu
\dfrac{\p \tilde u}{\p \bar x}
\right)
+\dfrac{\p }{\p \bar x}
\left(
\tilde \mu\dfrac{\p \tilde u}{\p y}
\right)
\right\},
\end{split}
\end{equation}
\begin{equation}
\begin{split}
&
\tilde \rho\dfrac{\p \tilde w}{\p \bar t}
+\tilde \rho \tilde u\dfrac{\p \tilde w}{\p \bar x}
+\tilde \rho \tilde v\dfrac{1}{\chi}\dfrac{\p \tilde w}{\p y}
+\tilde \rho \tilde w\dfrac{\p \tilde w}{\p z}
=
\\ &
\dfrac{1}{\chi^2}
\left\{ -\dfrac{\p \tilde p}{\p z}
+
\dfrac{\p }{\p z}
\left[
\dfrac{2}{3}\tilde \mu
\left(
2\dfrac{\p \tilde w}{\p z}
-
\dfrac{1}{\chi}
\dfrac{\p \tilde v}{\p y}
\right)
\right]
+
\dfrac{\p }{\p y}
\left[
\tilde \mu
\left(
\dfrac{1}{\chi}\dfrac{\p \tilde v}{\p z}
+\dfrac{\p \tilde w}{\p y}
\right)
\right]
-
\dfrac{\p }{\p z}
\left(
\dfrac{2}{3}\tilde \mu
\dfrac{\p \tilde u}{\p \bar x}
\right)
+
\dfrac{\p }{\p \bar x}
\left(
\tilde \mu\dfrac{\p \tilde u}{\p z}
\right)
\right\},
\end{split}
\end{equation}
\begin{equation}
\begin{split}
&
\tilde \rho\dfrac{\p \tilde T}{\p \bar t}
+\tilde \rho \tilde u\dfrac{\p \tilde T}{\p \bar x}
+\tilde \rho \tilde v\dfrac{1}{\chi}\dfrac{\p \tilde T}{\p y}
+\tilde \rho \tilde w\dfrac{\p \tilde T}{\p z}
=
\\ &
\dfrac{1}{\chi^2}
\left\{
\dfrac{1}{Pr}
\left[
\dfrac{\p }{\p y}
\left(
\tilde \mu\dfrac{\p \tilde T}{\p y}
\right)
+\dfrac{\p }{\p z}
\left(
\tilde \mu\dfrac{\p \tilde T}{\p z}
\right)
\right]
+
(\gamma-1)\mathcal{M}_\infty^2\tilde \mu
\left[
\left(
\dfrac{\p \tilde u}{\p y}
\right)^2
+
\left(
\dfrac{\p \tilde u}{\p z}
\right)^2
\right]
\right\},
\label{eqn:nubr-te}
\end{split}
\end{equation}
where $\chi=\sqrt{k_xR_\Lambda}$ and $\tilde \mu$ is the dynamic viscosity. The flow field is decomposed as the sum of the laminar flow and the perturbation,
\begin{eqnarray}
\left\{ \tilde{u}, \tilde{v},\tilde{w},\tilde {p}, \tilde{T} \right\}=\left\{U(\bar x, \eta),V(\bar x, \eta),0,\dfrac{1}{\gamma \mathcal{M}^2_\infty},T(\bar x, \eta)\right\} 
+r_t\left\{\bar u,\bar v, \bar w, \bar p, \bar \tau\right\}(\bar x, \eta,z,\bar t), \quad
\label{eqn:field-all}
\end{eqnarray}
where $r_t=\epsilon R_{\Lambda}= \mathcal{O}(1)$ is the turbulent Reynolds number, $$\eta=\sqrt{\dfrac{R_\Lambda}{2 x}}\int_0^y \dfrac{1}{T(\bar x,\check y)} {\rm d}{\check y},$$ and $\gamma$ is the ratio of specific heats. The perturbation in \eqref{eqn:field-all} represents the unsteady streaks, which are periodic in time with fundamental frequency $k_x$ (period $\mathcal{T} = 2\pi/k_x$). The streaky boundary layer \eqref{eqn:field-all} is the base flow for our secondary-instability analysis. The first term in \eqref{eqn:field-all} is the compressible Blasius boundary layer
\begin{equation}
{U}={F'(\eta), \quad V= \dfrac{T(\eta_c F'-F)}{\sqrt{2\bar x}}} \quad {\rm and}\quad T=T(\eta),
\end{equation}
where 
$$\eta_c=\dfrac{1}{T}\int_0^\eta T(\check \eta) {\rm d}{\check \eta},$$
and the prime denotes differentiation with respect to the similarity variable $\eta$. The compressible Blasius functions $F(\eta)$ and $T(\eta)$ satisfy the boundary-value problem
\begin{equation}
\left.\begin{array}{c}
\left(\dfrac{\mu F''}{T}\right)'+FF'' =0, \\
\left( \dfrac{\mu T'}{T} \right)'+PrFT'+ \mu(\gamma -1)Pr \mathcal{M}^2_\infty \dfrac{(F'')^2}{T}=0, \\
F=F'=0, \quad T=T_w, \quad {\rm at } \quad \eta=0, \\
F'\rightarrow1, \quad T\rightarrow1,\quad {\rm as } \quad \eta \rightarrow\infty.
\end{array}\right\}
\label{eqn:com-blasius}
\end{equation}
The Prandtl number $Pr$ is assumed constant, $Pr=0.707$, and the dynamic viscosity is assumed to obey the power law $\tilde \mu(\tilde T)=\tilde T^{s}$ with $s=0.76.$ 
The power law for the viscosity matches the Sutherland law in the range of temperatures of interest, i.e. for $\mathcal{M}_\infty=5$ and intense wall cooling. The viscosity is modelled using the power law for analytical tractability within our asymptotic framework \citep{marensi2017nonlinear}. 

The boundary-layer disturbances are expanded into temporal and spanwise harmonics,
\begin{eqnarray}
\left\{\bar u ,\bar v,\bar w,\bar p, \bar \tau\right\}
=
\sum^{\infty}_{m,n=-\infty}
\Big{\{}\hat u_{m,n}(\bar x ,\eta ), \sqrt{2\bar x} \hat v_{m,n}(\bar x,\eta ), k_z^{-1} \hat w _{m,n}(\bar x,\eta ),
  \qquad
  \qquad
  \qquad
  \nonumber
  \\
\hat p_{m,n}(\bar x,\eta ),\hat \tau_{m,n}(\bar x,\eta ) \Big{\}}
\, {\rm e}^{i m \bar t+ i n k_z z}.
\quad
\label{eqn:fourier}
\end{eqnarray}
As the physical quantities are real, the Fourier coefficients are Hermitian, $\hat q_{-m,-n}=(\hat q_{m,n})_{\rm cc},$ where $\hat q$ stands for any of $\hat u,\hat v,\hat w,\hat p,\hat \tau.$ Substituting expressions \eqref{eqn:field-all} and \eqref{eqn:fourier} into the nonlinear boundary-region equations \eqref{eqn:nubr-ta}–\eqref{eqn:nubr-te} yields the governing equations for the harmonics, given by (2.23)--(2.27) in \cite{marensi2017nonlinear}.

Matching the boundary-region solution with the outer solution yields the outer boundary conditions
\begin{linenomath*}
\begin{equation}
\left\{\hat{u}_{m,n},\hat{v}_{m,n},\hat{w}_{m,n},\hat{p}_{m,n},\hat{\tau}_{m,n}\right\}
\rightarrow \left\{0,
\dfrac{\kappa_z}{\sqrt{2\bar{x}}}v^{\dagger}_{m,n}, \kappa_z^2
w^{\dagger}_{m,n}, \dfrac{\epsilon}{k_x} p^{\dagger}_{m,n}, 0
\right\}  \ \ \text{as} \ \ \eta \rightarrow \infty,\label{eqn:BC}
\end{equation}
\end{linenomath*}
where $\kappa_z$=$k_z/\sqrt{k_xR_{\Lambda}}$ and $v^{\dagger}_{m,n}, w^{\dagger}_{m,n}$, $p^{\dagger}_{m,n}$ are given by (2.76) in \citet{marensi2017nonlinear}. These outer boundary conditions are derived by accounting for the mutual interactions of the FVD and the boundary layer. The present free-stream vortical disturbance consisting of a pair of oblique waves generates nonlinear terms in the boundary-region equations \eqref{eqn:nubr-ta}–\eqref{eqn:nubr-te} while providing a mathematically tractable description in the far field. The nonlinear terms cancel for this pair of oblique forcing modes, so that the governing equation in the far field is linear, as shown in equation~(2.69) of \cite{marensi2017nonlinear}. The displacement effects of the spanwise-modulated boundary layer induce nonlinear harmonics there, transmitted from the boundary layer via the displacement effect, as shown in equations~(2.74)--(2.76) of \cite{marensi2017nonlinear}. At the wall, the no-slip boundary condition for the velocity components is imposed and the temperature fluctuations vanish.

The velocity and temperature fluctuations induced near the leading edge arise from the entrainment of the FVD into the boundary layer. These perturbations are of small amplitude and therefore evolve linearly in this region. The upstream (initial) conditions imposed near the leading edge for the forced modes $(m,n)=(1,\pm 1)$ are thus the same as those in \citet{ricco2007response} and are given by (A2)--(A6) in \cite{xu2024excitation}. The resulting initial-boundary-value problem governs the excitation and nonlinear evolution of supersonic streaks in the presence of FVD for $r_t=\mathcal{O}(1)$ and $\mathcal{M}_{\infty}=\mathcal{O}(1).$ It is worth noting that the popular approach of specifying FVD and inlet conditions in terms of the continuous spectra of the Orr-Sommerfeld and Squire equations is inappropriate, as demonstrated by \cite{dong-wu-2013} and \cite{wu-dong-2016}. The flaw in that framework lies in the entrainment of FVD being influenced at leading order by nonparallelism. The latter is neglected in the continuous spectrum approach, and, as a result, the perturbations in the free stream and the boundary layer exhibit several physically unacceptable and mathematically troublesome behaviours.

\subsection{Linear instability of compressible streaks}
A biglobal instability analysis is used to study the stability of the streaky boundary layers induced by FVD. This analysis is local in the streamwise direction and instantaneous in time, but global in the wall-normal and spanwise directions. The two-dimensional nature of the eigenvalue problem renders this approach necessary as both the base flow and the disturbance flow vary in the wall-normal and spanwise directions \citep{theofilis2011global,xu_zhang_wu_2017,xu2024excitation}. As the unsteady streaks vary slowly with $\bar{x}$ and $\bar{t}$, the dependence on these two variables can be treated as parametric when the short-wavelength and high-frequency instability is considered. The local parallel-flow and quasi-steady approximations adopted herein are consistent with our asymptotic framework. The problem should therefore not be interpreted as a global stability analysis of the complete spatially developing flow in the sense of \cite{monkewitz1990local}. A fully global treatment of spanwise-periodic configurations could be formulated using approaches such as that of \cite{SchmidDePandoPeake2017}.
  
In view of the time-periodic nature of the streaks, temporal Floquet theory provides the rigorous framework for linear stability analysis when the base flow is strictly periodic in time. This theory describes the global instability over the entire temporal period, as in \citet{hack2014streak}, but leads to an expensive computational task. The instantaneous-instability analysis adopted here is a more efficient timewise-local approximation, in which the base flow is frozen at each phase and its instability characteristics are evaluated separately. This approach has been employed in our previous studies to capture the phase-dependent instability and its rapid amplification \citep{ricco2011evolution,xu_zhang_wu_2017,xu2024excitation}. The resulting behaviour is consistent with the intermittent bursts of high-frequency disturbances observed in streaky boundary layers and simple time-periodic flows \citep{luo2010linear}. The rapid local growth may drive the secondary disturbances to finite amplitude and thus to the nonlinear regime before one period of the base-flow evolution is completed. The linear secondary-instability analysis becomes invalid beyond this point.

A linear temporal instability mode is sought in the form
\begin{eqnarray}
\left\{u_s,v_s,w_s,p_s,\tau_s\right\}(x,y,z,t)=\left\{\hat u_s,\hat v_s,\hat w_s,\hat p_s,\hat \tau_s\right\}(y,z)
e^{i(\alpha x-\omega t)}+{\rm c.c.},
\label{eqn:sec}
\end{eqnarray}
where $\alpha \in \mathbb{R}$ is the streamwise wavenumber and $\omega \in \mathbb{C}$ is the frequency of the disturbance. The subscripts $r$ and $i$ henceforth indicate the real and imaginary parts, respectively. Substitution of \eqref{eqn:field-all} and \eqref{eqn:sec} into the compressible Navier-Stokes equations leads to the temporal stability equations. 
As the base flow is spanwise-periodic, $\left\{\hat u_s,\hat v_s,\hat w_s,\hat p_s,\hat \tau_s\right\}$ can be expressed using a Floquet expansion in the spanwise direction as
\begin{eqnarray}
\hat q_s=e^{{{ i} \xi \beta z}} \sum_{k=-\infty}^{\infty}\hat \phi_{s,k}(y) e^{{{ i} k\beta z}},
\end{eqnarray}
where $\beta$ is the spanwise wavenumber and $0\le \xi \le 1/2.$ Fundamental modes ($\xi = 0$), subharmonic modes ($\xi = 1/2$) and detuned modes ($0 < \xi < 1/2$) have different spanwise wavelengths, but they are all part of the same branch of instability. Consistent with previous studies \citep{ren2015secondary,paredes2019instability}, we have verified that the growth rate of the modes is only weakly dependent on the Floquet parameter, as shown in Appendix \ref{app:check}.

The shape functions $\left\{\hat u_s,\hat v_s,\hat w_s,\hat p_s,\hat \tau_s\right\}$ are governed by a system of partial differential equations, supplemented by homogeneous boundary conditions, $\left\{\hat u_s,\hat v_s,\hat w_s,\hat \tau_s\right\}=0$ at $y=0$ and $\left\{\hat u_s,\hat v_s,\hat w_s,\hat \tau_s\right\}\rightarrow 0$ as $y\rightarrow \infty$ for non-radiating modes. For radiating modes, the outer boundary conditions correspond to the modal disturbances attenuating exponentially at a small rate proportional to $\omega_i$ for temporal instability or to an outgoing wave in the wall-normal direction. In the far field, the equations reduce to a single equation for the pressure disturbance $\hat p_s(y,z),$ 
\begin{eqnarray}
\left[-\alpha^2\left(1-\mathcal{M}^2\right)+\frac{\p^2}{\p y^2}+\frac{\p^2}{\p z^2} \right] \hat p_s=0,
\label{eqn:gov}
\end{eqnarray}
where $\mathcal{M}$ is the free-stream relative Mach number,
$\mathcal{M}=\mathcal{M}_\infty(1-c)$, and $c$ is the phase speed of the instability wave. When $|\mathcal{M}|>1,$ the mode is supersonic. In the far field, the solution is $\hat{p}_{s,k}=C_k {\rm e}^{-i\zeta_k y},$ where $C_k$ is a constant and $\zeta_k =\sqrt{\alpha^2(\mathcal{M}^2-1)-k^2 \beta^2}$ is taken to have a negative imaginary part so that the mode attenuates \citep{qin2024excitation,zhang2023generation}. With the appropriately chosen $\zeta_k$, the far-field condition is expressed as $$\dfrac{\mathrm{d}\hat p_{s,k}}{\mathrm{d} y}+ i \zeta_k \hat p_{s,k}=0.$$ The temperature and velocities are also oscillatory in the far field. A similar radiation boundary condition was also employed in \cite{zhu2022effects} and \cite{zheng2024effects}. 

Temporal stability theory offers reduced computational complexity and time. Its results can be converted to spatial instability via the Gaster transformation \citep{gaster1962note}, which was recently extended to a second-order formulation by \cite{xu2023spatial} (referred to as XLZW hereafter). For a physically realistic characterisation of the boundary-layer instability, we present both temporal and spatial instability results. The Gaster transformation relates the complex streamwise wavenumber $\tilde{\alpha}$ to the temporal eigenvalue $\omega$ for a real wavenumber $\alpha$ by 
\begin{equation}
\tilde{\alpha}
= \tilde{\alpha}_r + i\tilde{\alpha}_i 
=\alpha - \dfrac{i\omega_i}{C_g}
= \left( \alpha - \frac{\omega_i}{|C_g|} \sin \Theta \right) - i \left( \frac{\omega_i}{|C_g|} \cos \Theta \right),
\label{eq:first_order_secondary}
\end{equation}
where $C_g = \mathrm{d} \omega / \mathrm{d} \alpha \equiv |C_g| \mathrm{e}^{i \Theta}$ is the complex-valued group velocity. The unnecessary approximation on the real and imaginary parts of $C_g$, i.e. $C_{g,i}\ll C_{g,r}$, made by \cite{gaster1962note}, is avoided in expression \eqref{eq:first_order_secondary}. By retaining the quadratic terms in the Taylor expansion of the dispersion relation $\omega(\alpha),$ an improved relation for the spatial eigenvalue $\tilde{\alpha}$ is obtained:
\begin{equation}
\tilde{\alpha} = \tilde{\alpha}_r + i\tilde{\alpha}_i= \alpha+\frac{\mathrm{d}\omega }{\mathrm{d}\alpha} \left[ -1 + \sqrt{1 - 2 {i}\frac{\mathrm{d}^2\omega}{\mathrm{d}\alpha^2} \omega_i \bigg/ \left( \frac{\mathrm{d}{\omega}}{\mathrm{d}\alpha} \right)^2 } \right] \bigg/ \frac{\mathrm{d}^2{\omega}}{\mathrm{d}\alpha^2},
\label{eq:second_order_secondary}
\end{equation}
where the derivatives $\mathrm{d}\omega/\mathrm{d}\alpha$ and $\mathrm{d}^2\omega/\mathrm{d}\alpha^2$ are evaluated numerically by solving the temporal eigenvalue problem for small $\alpha$ perturbations along the imaginary axis. The second-order transformation \eqref{eq:second_order_secondary} was found to give more accurate growth rates than the original Gaster transformation.

\subsection{Numerical procedures} 
The nonlinear initial-boundary-value problem was discretised using a second-order backward finite-difference scheme in the streamwise direction and a second-order central finite-difference scheme in the wall-normal direction. The pressure was computed on a staggered grid to avoid decoupling. The resulting block tridiagonal system was solved via a standard block-elimination algorithm \citep{marensi2017nonlinear}, with the nonlinear terms being treated by a predictor--corrector scheme. An under-relaxation iteration was implemented to ensure convergence. Along the wall-normal direction, the domain extended to $\eta_{max} = 60$ and 2000 grid points were used. The step size in the marching direction was $\Delta \bar x = 0.001$. To capture nonlinear effects, it sufficed to retain 17 modes in both the frequency and spanwise wavenumber domains. 

The biglobal stability equations were discretised using a five-point finite-difference scheme with fourth-order accuracy along the wall-normal direction and a Fourier spectral expansion along the spanwise direction \citep{xu2024excitation}. The domain extended to $y=49$ along the wall-normal direction, i.e. five or six times the boundary-layer thickness. This domain size was sufficient to achieve converged results thanks to the boundary conditions employed for the non-radiating and radiating modes. At those wall-normal locations, the viscous effects are negligible and the base flow is spanwise uniform. Eight to ten spanwise Fourier modes were sufficient to compute the secondary-instability modes. The resulting linear generalised eigenvalue problem was solved by a Rayleigh quotient iteration method \citep{parlett1998symmetric}.

\section{Results}
\label{sec:results}
The excitation and evolution of supersonic streaks over cooled walls are discussed in \S\ref{sec:streaks} and the instability of these streaks is presented in \S\ref{sec:first-family} and \S\ref{sec:second-family}. Two families of streak-instability modes are identified. The first family refers to the modified Mack modes, which can be radiating or non-radiating. They reduce to the classical Mack second modes in the limit of vanishingly small streak amplitude, and are substantially altered when the streak amplitude is increased, as discussed in \S\ref{sec:first-family}. The second family consists of three-dimensional secondary-instability modes that exist only in the presence of streaks with sufficiently large amplitude, as demonstrated in \S\ref{sec:second-family}.

\subsection{Excitation and nonlinear evolution of high-supersonic streaks}
\label{sec:streaks}

Numerical calculations were performed for parameters pertinent to flight conditions in the high-supersonic regime with a free-stream Mach number $\mathcal{M}_\infty=5$ \citep{wilson1971evaluation,schneider1999flight,xu2023review}. For conditions representative of high-altitude flight between 20 and 25 km, we take the free-stream temperature and kinematic viscosity of air to be $T_\infty^*=218$ K and $\nu_\infty^*=2.67\times 10^{-4}$ m$^2$s$^{-1}$, respectively. The unit Reynolds number is {$Re_u={{U}_\infty^*} /{\nu_\infty^*}=5.54\times10^6$ m$^{-1}$ }(which corresponds to $U_\infty^*=1480$ ms$^{-1}$). The spanwise wavelength of the FVD that generates the streaks is {$\lambda_z^*= 11.4$ mm} and the frequency is {0.166 kHz} ($k_x=1.28 \times 10^{-3}$). It follows that the Reynolds number is $R_\Lambda=Re_{u}\lambda_z^*/2\pi=10029$ and $k_z=1.$ Without loss of generality, we set $\hat u^\infty_{z,\pm}=\pm 1$. The intensity $\epsilon$ of the FVD is related to the turbulence level $Tu$ by $Tu=  2\epsilon\left(\hat u_{x,+}^{\infty 2}+\hat u_{x,-}^{\infty 2}\right)^{1/2}$ \citep{marensi2017nonlinear}. In high-supersonic and hypersonic flights, the typical turbulence level for quiescent atmosphere is about $0.05\%$ or even lower \citep{Zanin1985}. We, however, consider a broader range of FVD intensities to encompass different flight conditions, such as those close to the inversion layer or within clouds, where the turbulence level can reach $0.15\%$. The reader is referred to \citet{marensi2017nonlinear} for details on the parameter selection.

\begin{figure}
\centering
\subfigure{
\includegraphics[width=0.42\textwidth]{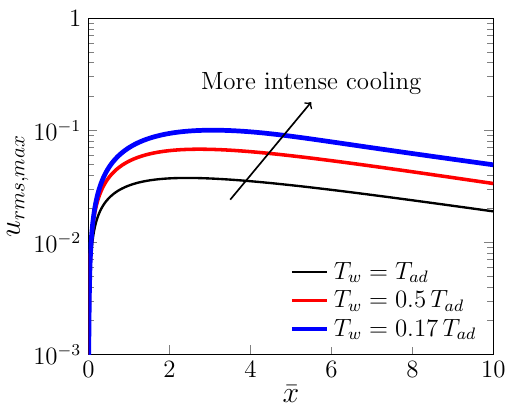}
\put(-155,117){($a$)}
}
\subfigure{
\includegraphics[width=0.42\textwidth]{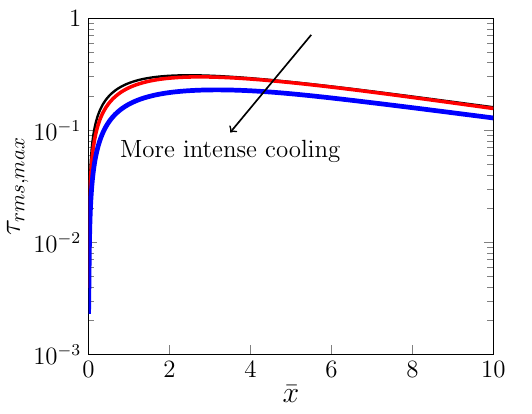}
\put(-155,117){($b$)}
\psfrag{$10^0$}{$1$}
}
\vspace{-0.4cm}
\subfigure{
\includegraphics[width=0.33\textwidth]{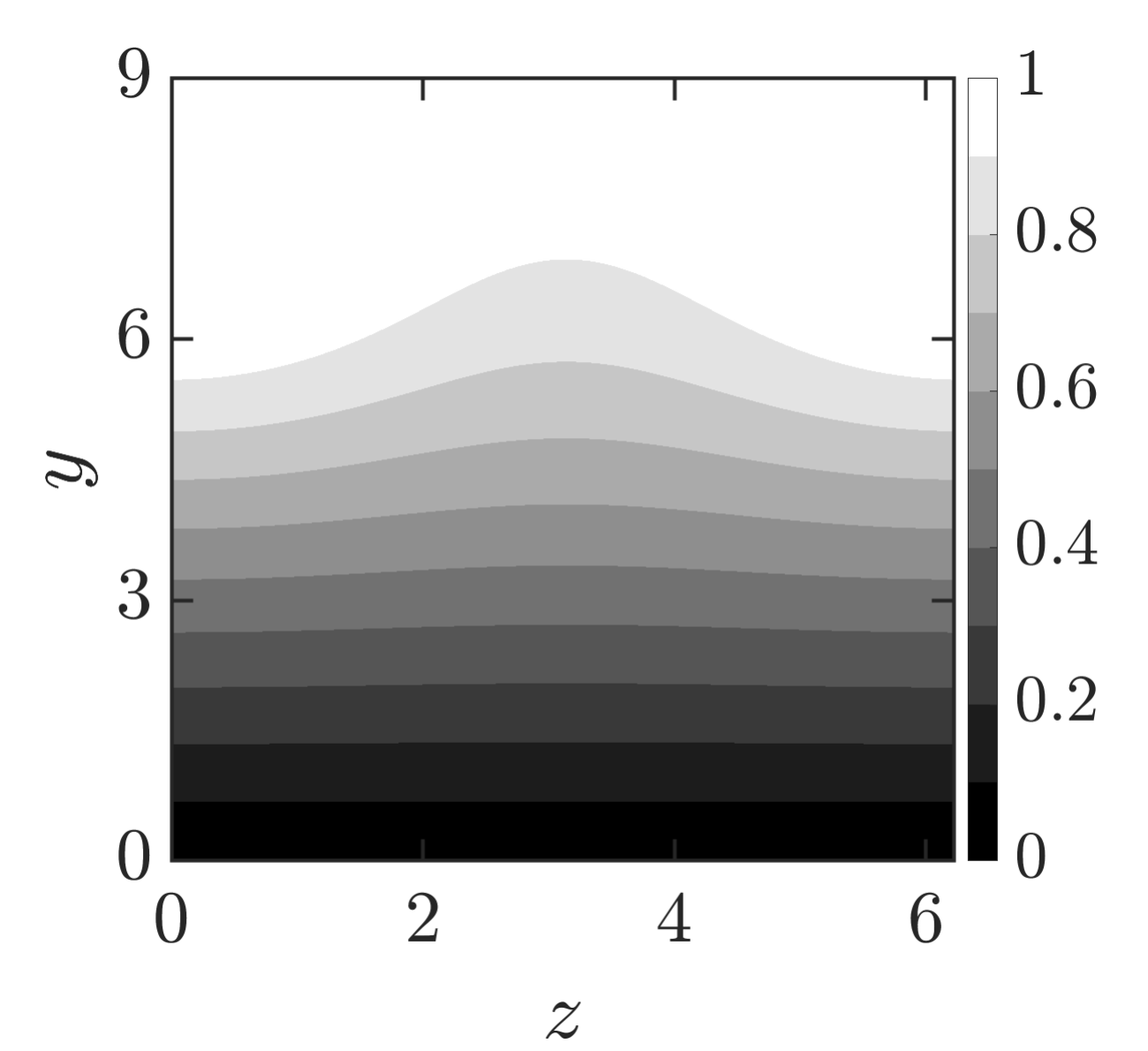}
\put(-85,112){$T_w=T_{ad}$}
\put(-100,93){\color{black}ISV}
 \put(-128,104){{($c$)}}
 \put(-120,62){\makebox(0,0){\colorbox{white}{$y$}}}
\put(-66,6){\makebox(0,0){\colorbox{white}{$z$}}}
}
\hspace{-0.3cm}
\subfigure{
\includegraphics[width=0.33\textwidth]{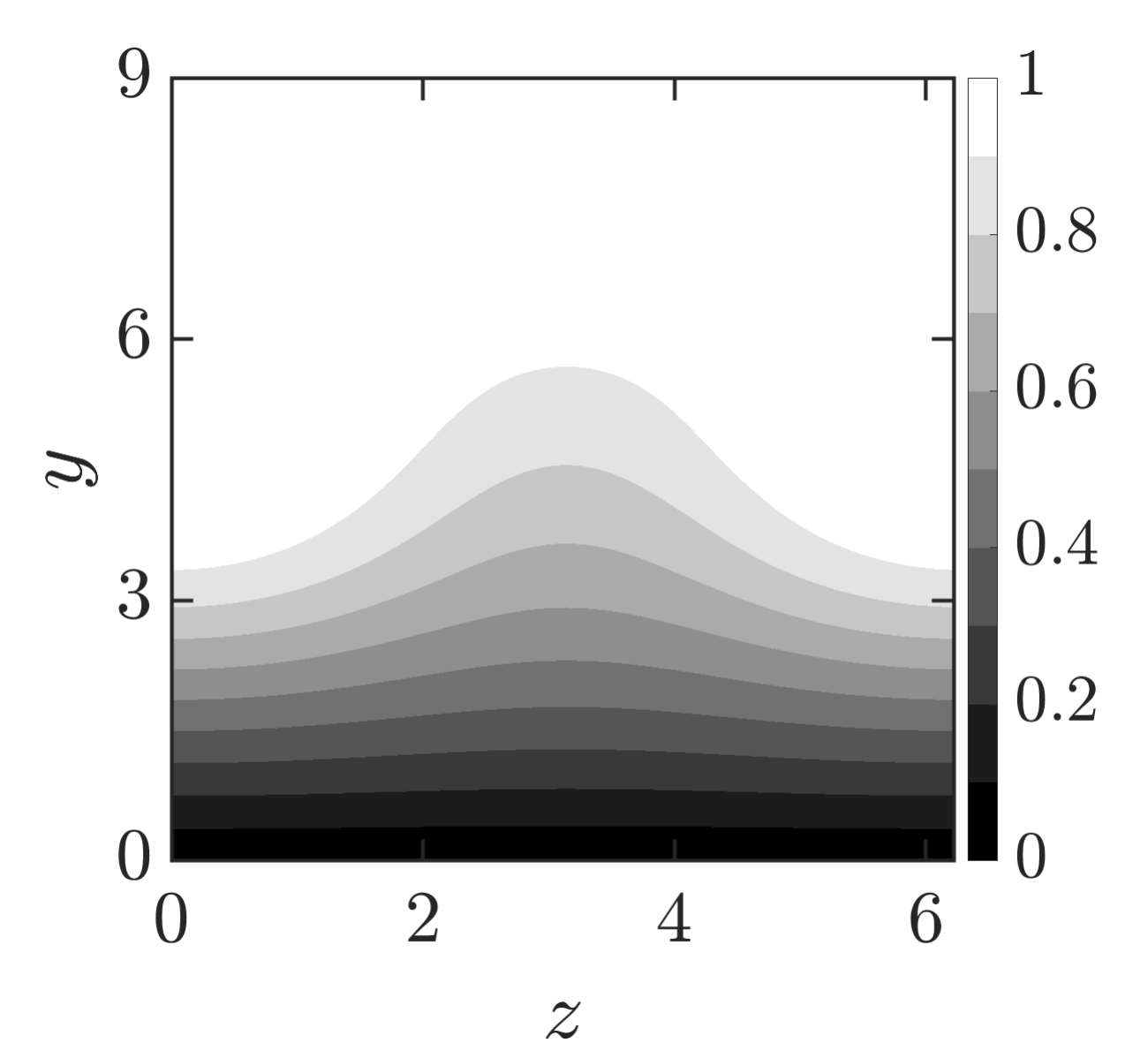}
\put(-85,112){$T_w=0.5T_{ad}$}
\put(-100,93){\color{black}ISV}
\put(-128,104){{($d$)}}
\put(-120,62){\makebox(0,0){\colorbox{white}{$y$}}}
\put(-66,6){\makebox(0,0){\colorbox{white}{$z$}}}
}\hspace{-0.5cm}
\subfigure{
\begin{tikzpicture}[baseline={([yshift=3pt] current bounding box.south)}]
\node(a){
\includegraphics[width=0.33\textwidth]{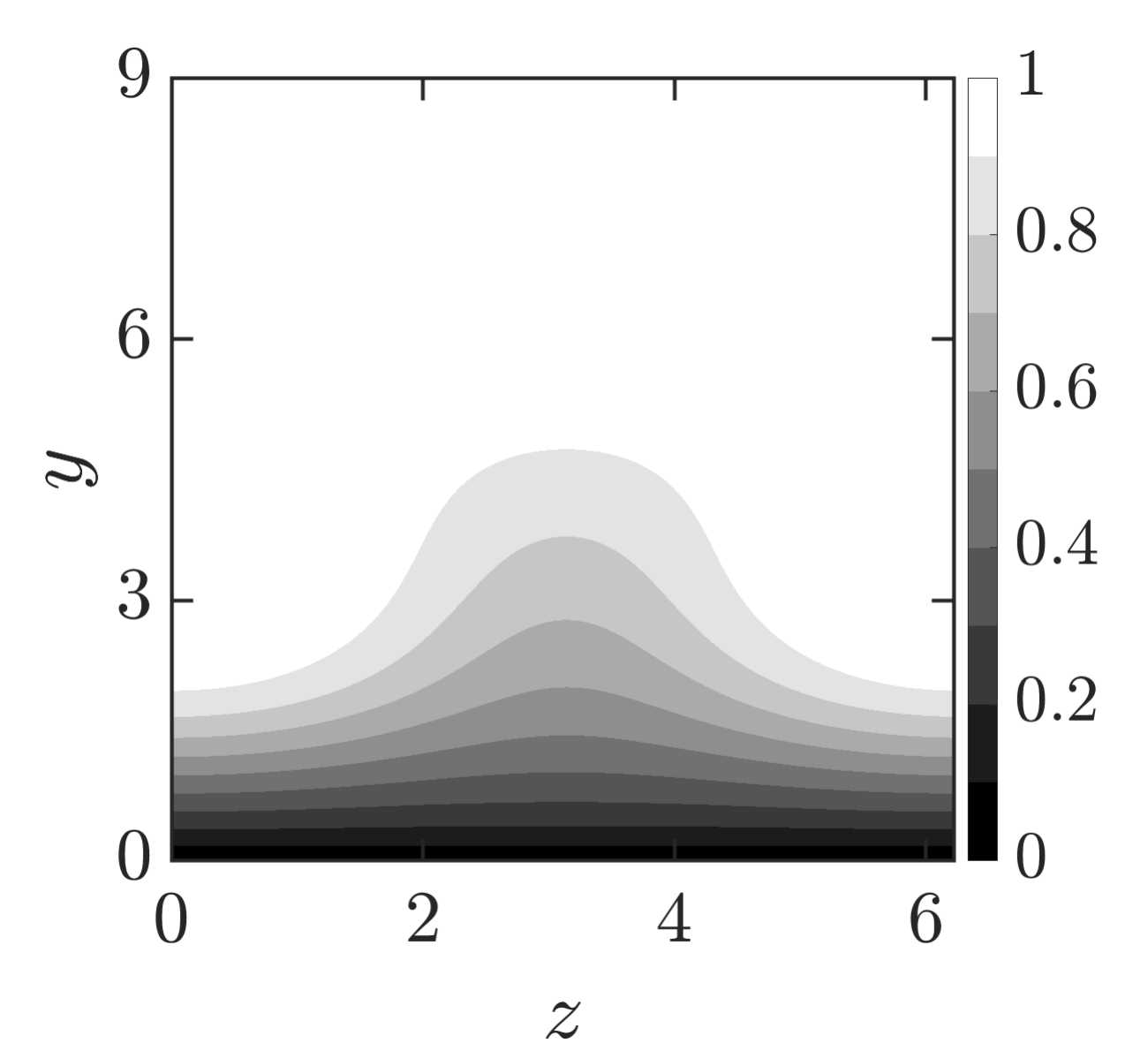}};
\node at(a.center)[draw, black,line width=.6pt,ellipse, minimum width=35, minimum height=25pt,rotate=-0,yshift=-8pt,xshift=-2pt]{Peak};
\draw[black, line width=.6pt] (a.center) ++(-44pt,-28pt) arc[green,start angle=-90, end angle=90, x radius=25pt, y radius=10pt];
\node at (a.center) [ black,yshift=-18pt,xshift=-32pt] {Valley};
\draw[black, line width=.6pt] (a.center) ++(42pt,-8pt) arc[start angle=90, end angle=270, x radius=27pt, y radius=10pt];
\node at (a.center) [ black,yshift=-18pt,xshift=30pt] {Valley};
\end{tikzpicture}
\put(-130,104){{($e$)}}
\put(-92,112){$T_w=0.17 T_{ad}$}
\put(-104,93){\color{black}ISV}
\put(-120,62){\makebox(0,0){\colorbox{white}{$y$}}}
\put(-66,6){\makebox(0,0){\colorbox{white}{$z$}}}
}
\vspace{-0.85cm}
\subfigure{
\includegraphics[width=0.33\textwidth]{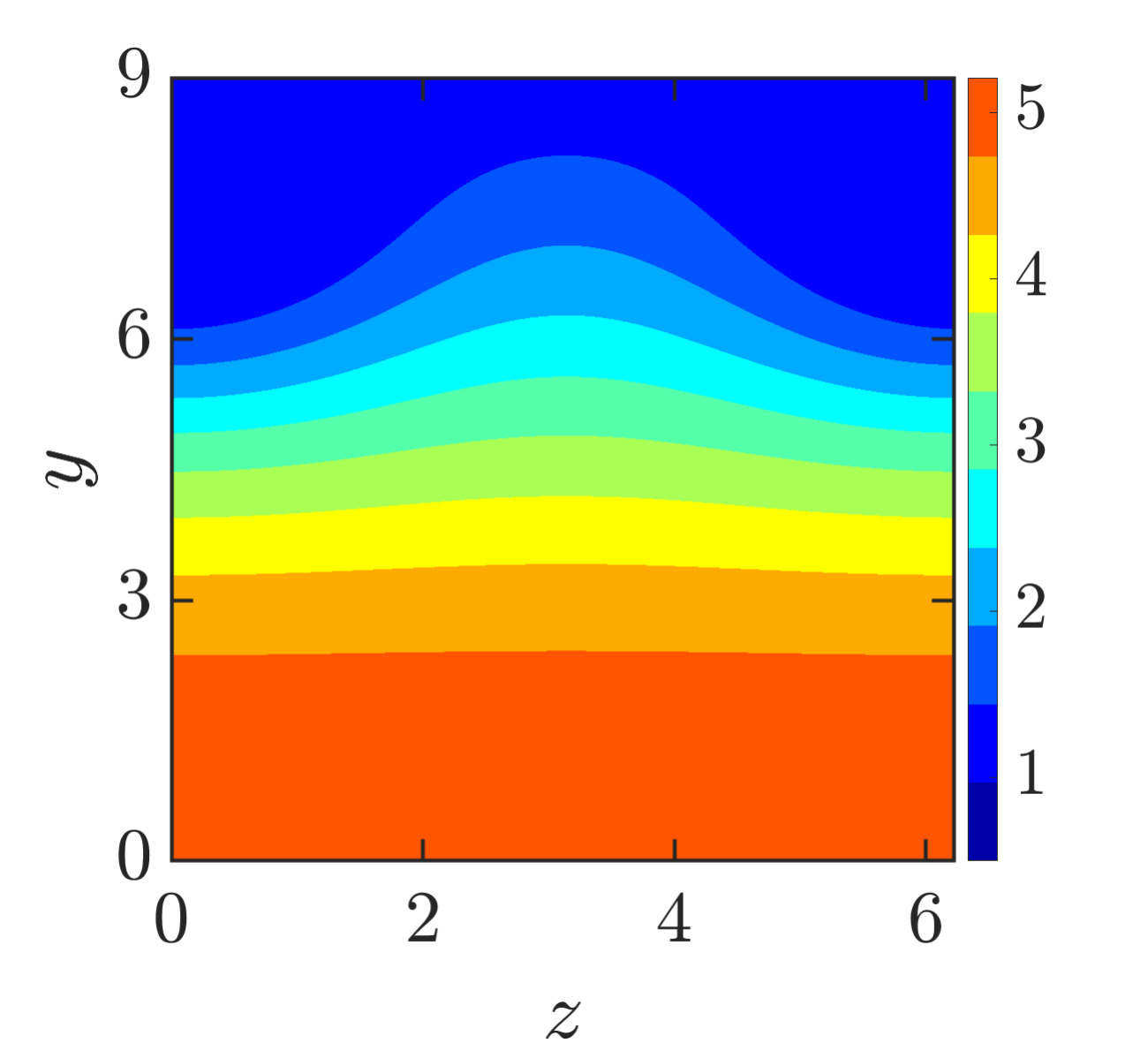}
 \put(-128,104){{($f$)}}
 \put(-100,93){\color{white}IT}
 \put(-120,62){\makebox(0,0){\colorbox{white}{$y$}}}
\put(-66,6){\makebox(0,0){\colorbox{white}{$z$}}}

}\hspace{-0.2cm}
\subfigure{
\includegraphics[width=0.33\textwidth]{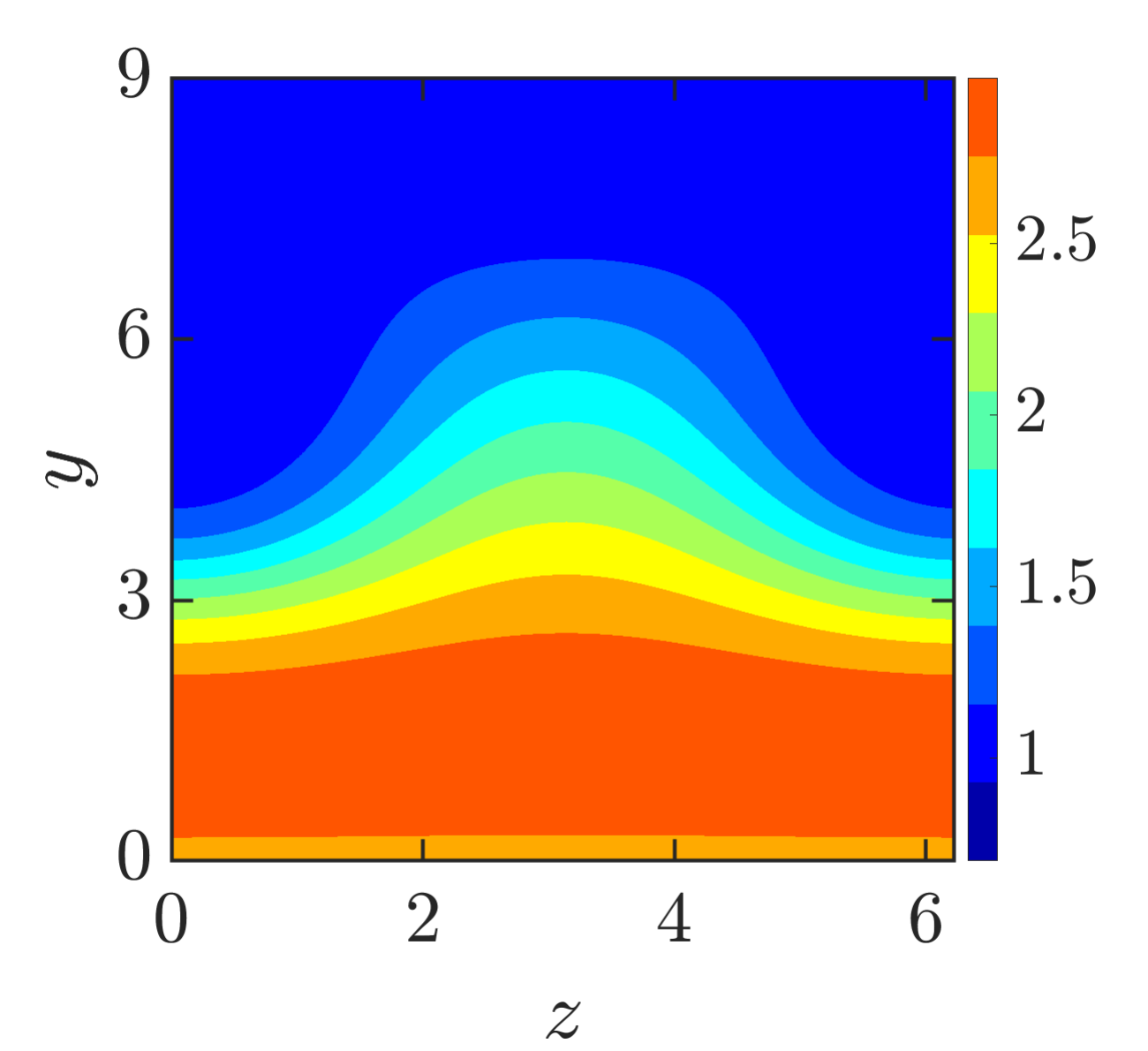}
 \put(-128,104){{($g$)}}
  \put(-100,93){\color{white}IT}
  \put(-120,62){\makebox(0,0){\colorbox{white}{$y$}}}
\put(-66,6){\makebox(0,0){\colorbox{white}{$z$}}}
}\hspace{-0.35cm}
\subfigure{
\begin{tikzpicture}[baseline={([yshift=3pt] current bounding box.south)}]
\node(a){
\includegraphics[width=0.33\textwidth]{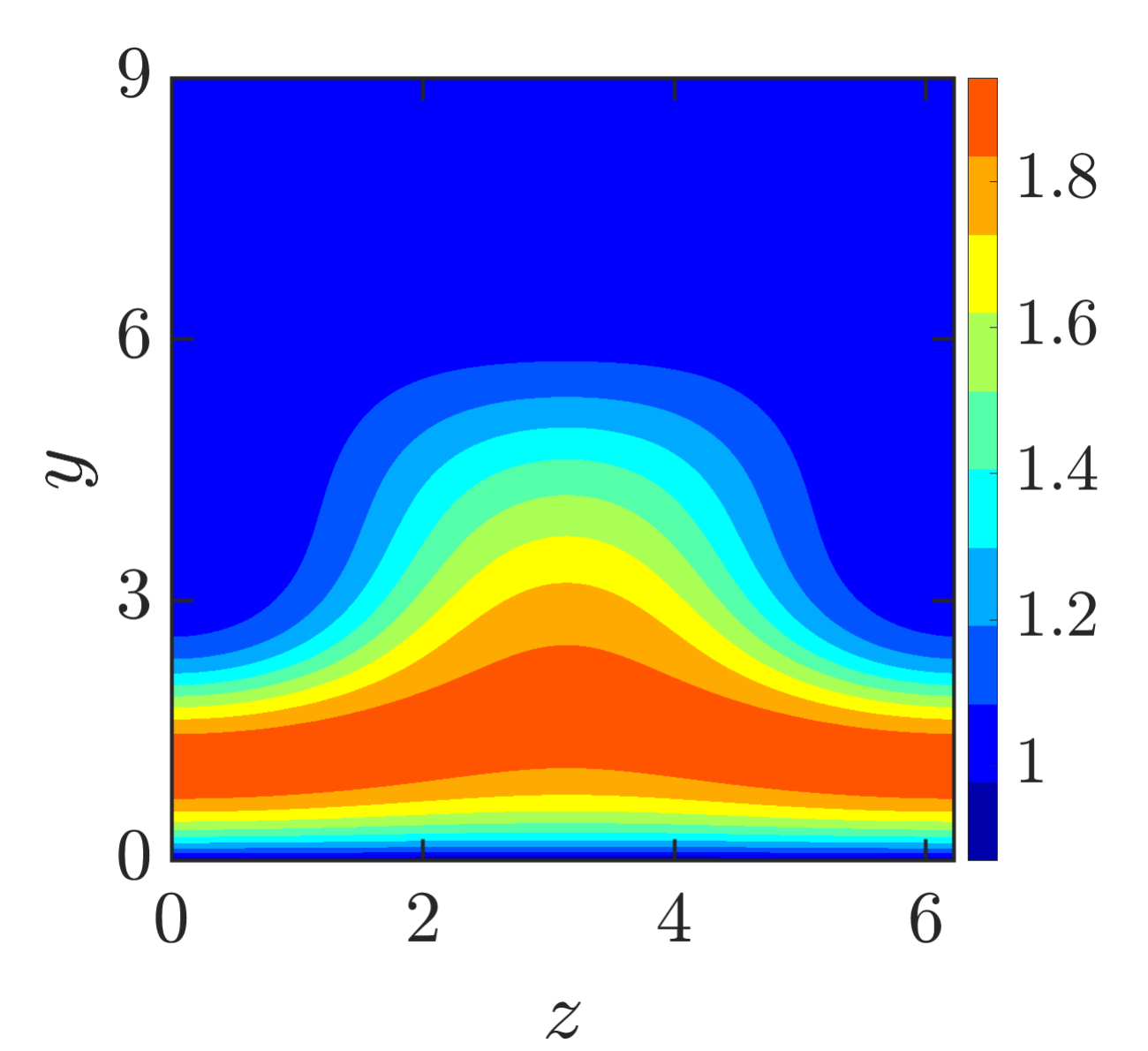}};
\node at(a.center)[draw, black,line width=.6pt,ellipse, minimum width=1pt, minimum height=8pt,rotate=-0,yshift=-18pt, inner sep=1pt, font=\small,  text width=50pt, align=center]{Internal heating band};

\end{tikzpicture}
 \put(-104,93){\color{white}IT}
 \put(-120,62){\makebox(0,0){\colorbox{white}{$y$}}}
\put(-66,6){\makebox(0,0){\colorbox{white}{$z$}}}
\put(-130,104){($h$)}
}
\vspace{5mm}
\captionsetup{justification=raggedright}
\caption{Evolution of maximum r.m.s. of ($a$) streamwise velocity and ($b$) temperature disturbances at three different wall temperatures. Contours of ($c$-$e$) instantaneous streamwise velocity (ISV) and ($f$-$h$) instantaneous temperature (IT) in the $y-z$ plane at $\bar x=4.0$. The parameters are $r_t=3.6,$ $R_\Lambda=10029$ and $\mathcal{M}_\infty=5.$  The phase is $\phi \equiv k_xt = 7\pi/16$.
}
\label{fig:utrms-tw}
\end{figure}

Before examining the instability, we first study how the wall temperature and the FVD level influence the velocity and thermal streaks. The downstream development of the streaks is quantified by the maximum root mean square (r.m.s.) of the streamwise velocity and temperature. The wall temperature takes the adiabatic-wall value, \(T_w = T_{{ad}}\), and two cooled-wall values, \(T_w = 0.5T_{{ad}}\) and \(T_w =0.17T_{{ad}}\), typical for high-speed flight conditions. Figure \ref{fig:utrms-tw}($a$,$b$) shows the effect of the wall temperature on the evolution of the supersonic streaks. The maximum r.m.s. of the velocity is smaller than that of the temperature. The cooled wall enhances the velocity disturbances but inhibits the temperature disturbances. The latter trend was also observed by \citet{marensi2017nonlinear} at $\mathcal{M}_\infty=3$.
Figure \ref{fig:utrms-tw}($c$-$h$) shows the effect of the wall temperature on the instantaneous streamwise velocity and the instantaneous temperature of the streaks at a streamwise location and temporal phase at which the streaks have reached a large amplitude and are strongly distorted. The cooled wall causes the streaks to move closer to the wall. The shape of the temperature contour is slightly more modified than that of the velocity, leading to the formation of saturated bell-shaped thermal streaks (figure \ref{fig:utrms-tw}$h$). Temperature contours were not reported by \cite{marensi2017nonlinear} and \cite{paredes2019instability}, despite thermal effects being dominant in high-supersonic and hypersonic boundary layers \citep*{huang2021early}. The most severely heated region, due to aerodynamic heating, is near the wall for \(T_w = T_{ad}\). It becomes thinner and shifts upward toward the core of the boundary layer for a wall temperature typical of flight conditions (\(T_w =0.17T_{ad}\), figure \ref{fig:utrms-tw}$h$). Consequently, an internal heating band characterised by intense temperature gradients forms in the bulk of the cooled-wall boundary layer. The high temperature within the heating band significantly modifies the local velocity profile because the thermal expansion reduces density and enhances shear. As shown in figure \ref{fig:utrms-tw}($e$), the cooled bell-shaped streaks exhibit two distinct regions. The streak peaks correspond to low-speed, high-temperature zones where the local momentum transport is suppressed while the local heat transfer is enhanced. The valleys, instead, correspond to high-speed, low-temperature regions where the boundary layer becomes thinner.

\begin{figure}
\centering
\subfigure{
\includegraphics[width=0.42\textwidth]{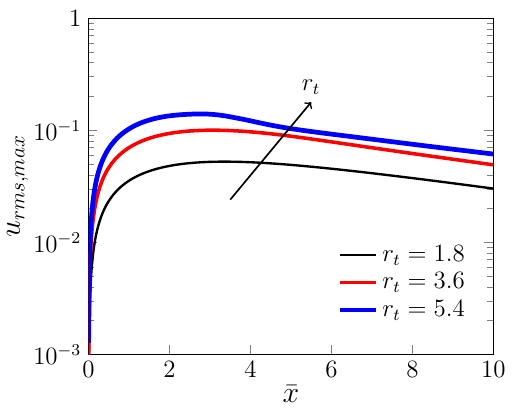}
\put(-155,117){($a$)}}
\subfigure{
\includegraphics[width=0.42\textwidth]{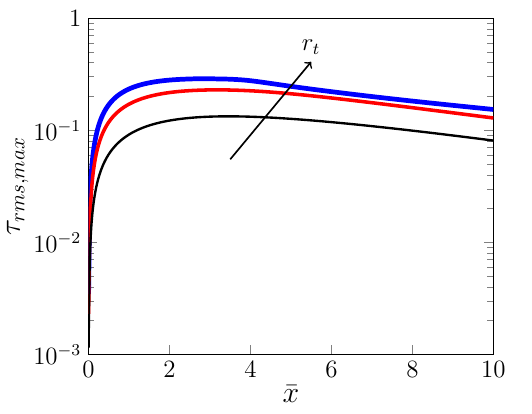}
\put(-155,117){($b$)}
}
\vspace{-0.4cm}
\subfigure{
\includegraphics[width=0.33\textwidth]{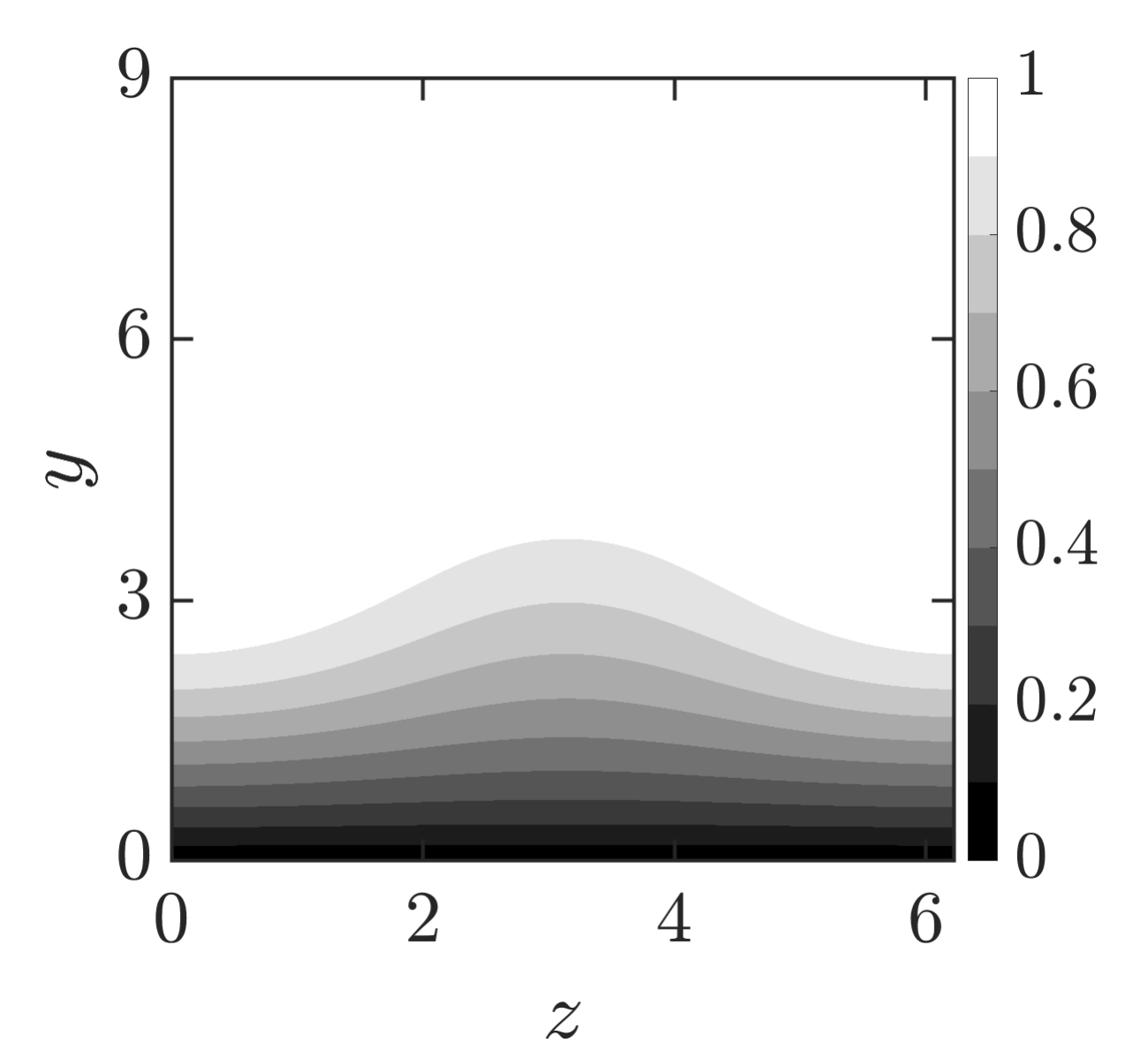}
\put(-128,104){{($c$)}}
\put(-81,112){$r_t=1.8$}
\put(-100,93){\color{black}ISV}
\put(-120,62){\makebox(0,0){\colorbox{white}{$y$}}}
\put(-66,6){\makebox(0,0){\colorbox{white}{$z$}}}
}\hspace{-0.3cm}
\subfigure{
\includegraphics[width=0.33\textwidth]{figs/figure1e.pdf}
\put(-81,112){$r_t=3.6$}
\put(-128,104){{($d$)}}
\put(-100,93){\color{black}ISV}
\put(-120,62){\makebox(0,0){\colorbox{white}{$y$}}}
\put(-66,6){\makebox(0,0){\colorbox{white}{$z$}}}
}\hspace{-0.3cm}
\subfigure{
\includegraphics[width=0.33\textwidth]{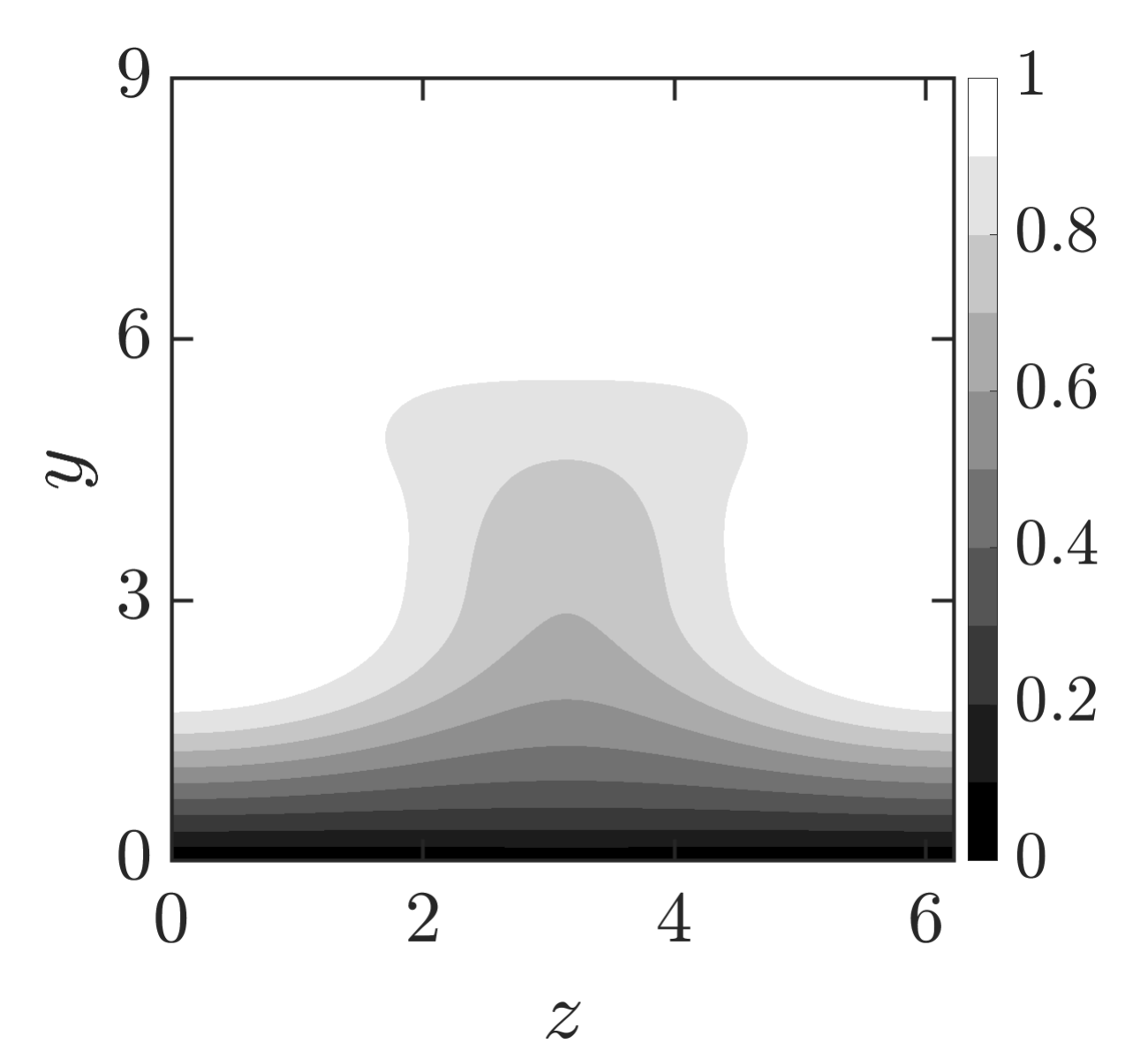}
\put(-81,112){$r_t=5.4$}
\put(-128,104){{($e$)}}
\put(-100,93){\color{black}ISV}
\put(-120,62){\makebox(0,0){\colorbox{white}{$y$}}}
\put(-66,6){\makebox(0,0){\colorbox{white}{$z$}}}
}
\vspace{-0.7cm}
\hspace{0.05cm}
\subfigure{
\includegraphics[width=0.33\textwidth]{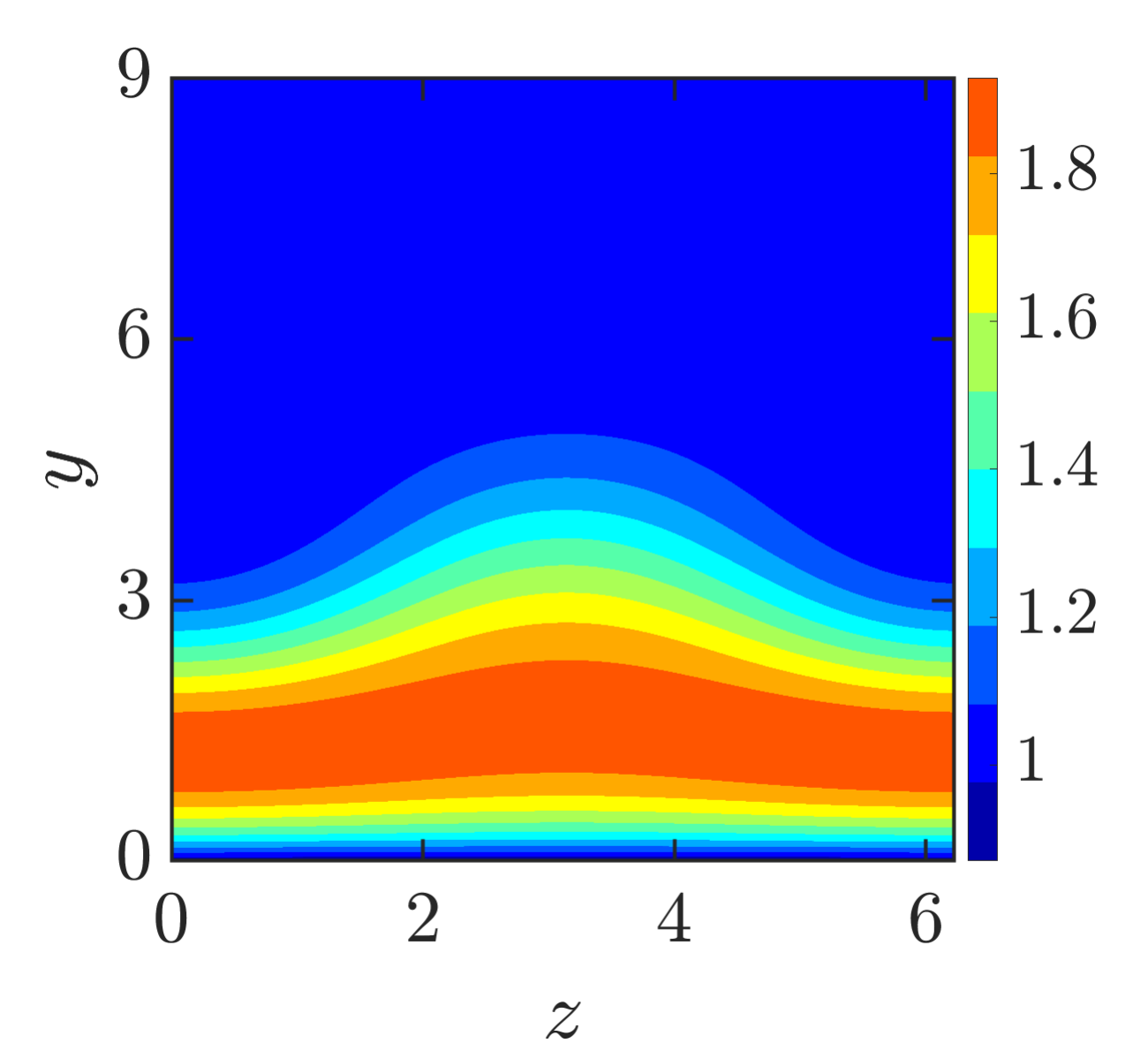}
\put(-128,104){{($f$)}}
 \put(-100,93){\color{white}IT}
 \put(-120,62){\makebox(0,0){\colorbox{white}{$y$}}}
\put(-66,6){\makebox(0,0){\colorbox{white}{$z$}}}
}\hspace{-0.3cm}
\subfigure{
\includegraphics[width=0.33\textwidth]{figs/figure1h.pdf}
 \put(-100,93){\color{white}IT}
\put(-128,104){{($g$)}}
\put(-120,62){\makebox(0,0){\colorbox{white}{$y$}}}
\put(-66,6){\makebox(0,0){\colorbox{white}{$z$}}}
}\hspace{-0.3cm}
\subfigure{
\includegraphics[width=0.33\textwidth]{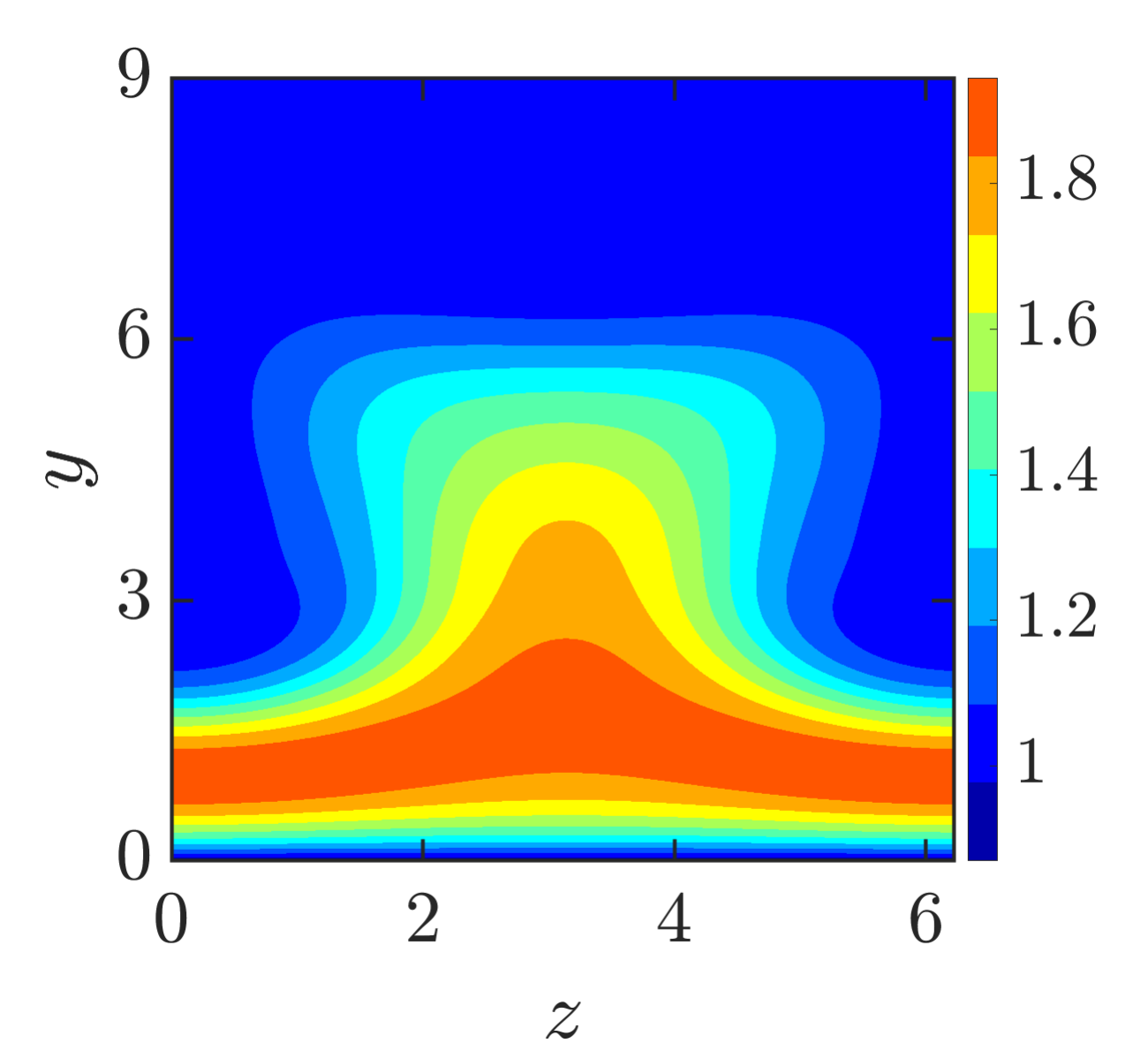}
\put(-128,104){{($h$)}}
\put(-100,93){\color{white}IT}
\put(-120,62){\makebox(0,0){\colorbox{white}{$y$}}}
\put(-66,6){\makebox(0,0){\colorbox{white}{$z$}}}
}
\vspace{5mm}
\captionsetup{justification=raggedright}
\caption{Evolution of maximum r.m.s. of ($a$) streamwise velocity and ($b$) temperature disturbances at three different turbulence levels. Contours of ($c$-$e$) instantaneous streamwise velocity (ISV) and ($f$-$h$) instantaneous temperature (IT) in the $y$-$z$ plane at $\bar x=4.0$. The parameters are $R_\Lambda=10029,$ $\mathcal{M}_\infty=5$ and $T_w=0.17T_{ad}.$ The phase is $\phi = 7\pi/16$.
}
\label{fig:utrms}
\end{figure}

In order to study the effect of the free-stream turbulence level, the turbulent Reynolds number is set to a value typical of flight conditions, \(r_t = 1.8\) \((Tu = 0.06\%)\), and to two higher values, \(r_t = 3.6\) \((Tu = 0.11\%)\) and \(r_t = 5.4\) \((Tu = 0.16\%\)). Figure \ref{fig:utrms}($a,b$) shows the evolution of the maximum velocity and temperature r.m.s. at different turbulent Reynolds numbers for the cooled-wall condition with $T_w=0.17T_{ad}$. Both the velocity and temperature disturbances are enhanced as $r_t$ is increased. Figure \ref{fig:utrms}($c$-$h$) displays contours of instantaneous streamwise velocity and temperature at different turbulent Reynolds numbers. The upper part of the cross-sectional shapes of the velocity and thermal disturbances becomes more distorted as $r_t$ increases. The internal heating bands, shown in figure \ref{fig:utrms}($f$-$h$), are weakly impacted.
All subsequent stability calculations use the instantaneous streaky base flow at $\bar{x}=4.0$ and $\phi=7\pi/16$, selected to examine flow conditions for which the streaks have reached their maximum amplitude and are highly distorted.

\subsection{Temporal and spatial modified Mack second modes}
\label{sec:first-family}
\begin{figure}
\centering
\subfigure{
\includegraphics[width=0.42\textwidth]{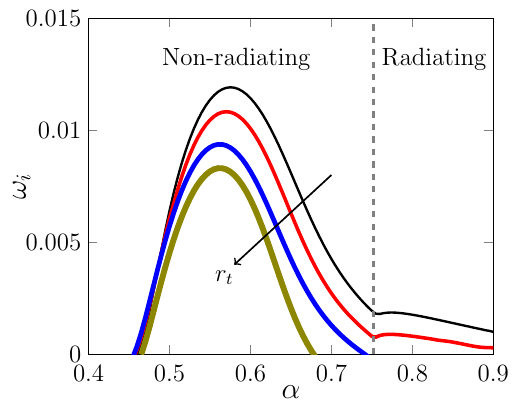}
\put(-165,118){($a$)}
}
\subfigure{
\includegraphics[width=0.42\textwidth]{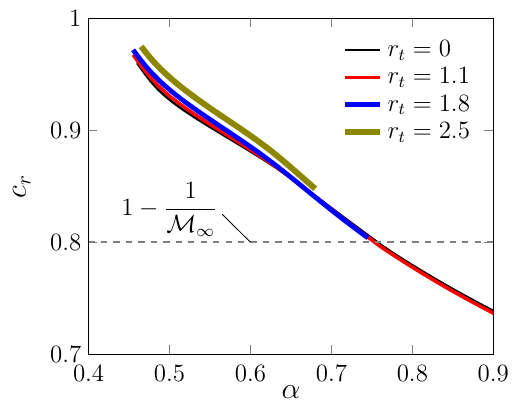}
\put(-165,118){($b$)}
}
\captionsetup{justification=raggedright}
\caption{Effect of the amplitude of high-supersonic streaks on Mack modes. ($a$) Temporal growth rate $\omega_i$; ($b$) phase velocity $c_r=\omega_r/\alpha$. The vertical and horizontal grey dashed lines correspond to the critical phase velocity, $c_{r,c} = 1 - 1/\mathcal{M}_\infty$, and separate the non-radiating and radiating modes. The  parameters are $R_\Lambda=10029,$ $\mathcal{M}_\infty=5$ and $T_w=0.17T_{ad}.$}
\label{2D-mod}
\end{figure}

For flat-plate boundary layers at $\mathcal{M}_\infty = 5$, two linearly unstable modes may exist: the first mode and the Mack second mode \citep{mack1984boundary}. We focus on the latter as it is the most amplified, while the first mode is stabilised by wall cooling. Figure \ref{2D-mod} shows the effect of the streaks on the Mack second modes. For a cooled-wall high-supersonic boundary layer, the growth rate curve of the Mack second mode displays two distinct unstable frequency ranges, separated by the critical phase velocity $c_{r,c} = 1 - 1/\mathcal{M}_\infty$ \citep{mack1984boundary}. As figure \ref{2D-mod}($a$) shows, the non-radiating Mack modes are stabilised by the streaks, an effect analogous to that reported by \citet{paredes2019instability} and \citet{caillaud2025} for optimal streaks. Here we report the effect of the streaks on the radiating modes for the first time: the streaks also inhibit the radiating modes, and, for $r_t \geq 1.8,$ no radiating modes are found. A radiating mode emits a highly directional Mach-wave beam as it undergoes amplification-attenuation near the neutral position. Its far-field acoustic pressure is explicitly related to the amplitude of the mode through a linear integral formula, given by (4.20) in \cite{wu2005mach}. This formula remains valid in the presence of streaks since the latter are trapped within the boundary layer. It follows that the streaks weaken (if $0<r_t<1.8$) and may even prevent (if $r_t>1.8$) emission of acoustic waves to the far field from the transitional boundary layer. A detailed mathematical explanation for the suppression of the radiating modes remains a subject for future investigation. Figure \ref{2D-mod}($b$) shows how the phase velocity $c_r$ changes with $\alpha$; the critical phase velocity $c_{r,c}$, separating non-radiating and radiating modes, is also displayed. The phase velocity of the non-radiating modes increases slightly with $r_t$, while that of the radiating modes remains unchanged. According to the far-field equation \eqref{eqn:gov}, the \(k\)th Fourier harmonic is purely radiating only if $c_i=0$, $c_r<c_{r,c}$ and $|k|\beta<\alpha\sqrt{\mathcal{M}_\infty^2(1-c_r)^2-1}$. For the present normalisation, $\beta=1$. The harmonic $k=0$ reaches the sonic threshold at  $\alpha=0.74$ and $c_{r,c}=0.8$, and becomes radiating for larger $\alpha$. The term $\alpha\sqrt{\mathcal{M}_\infty^2(1-c_r)^2-1}$ reaches a maximum of approximately 0.82, near $\alpha=0.9$ and $c_r=0.73$, therefore remaining below unity throughout the radiating interval. Only the spanwise-uniform harmonic $k=0$ can thus radiate, whereas all harmonics with $|k|\geq1$ are evanescent in the free stream. Hence, the radiation suppression by the streaks is attributed to the suppression of the growth of the radiating \(k=0\) harmonic.
\begin{figure}
\subfigure{
\includegraphics[width=0.24\textwidth]{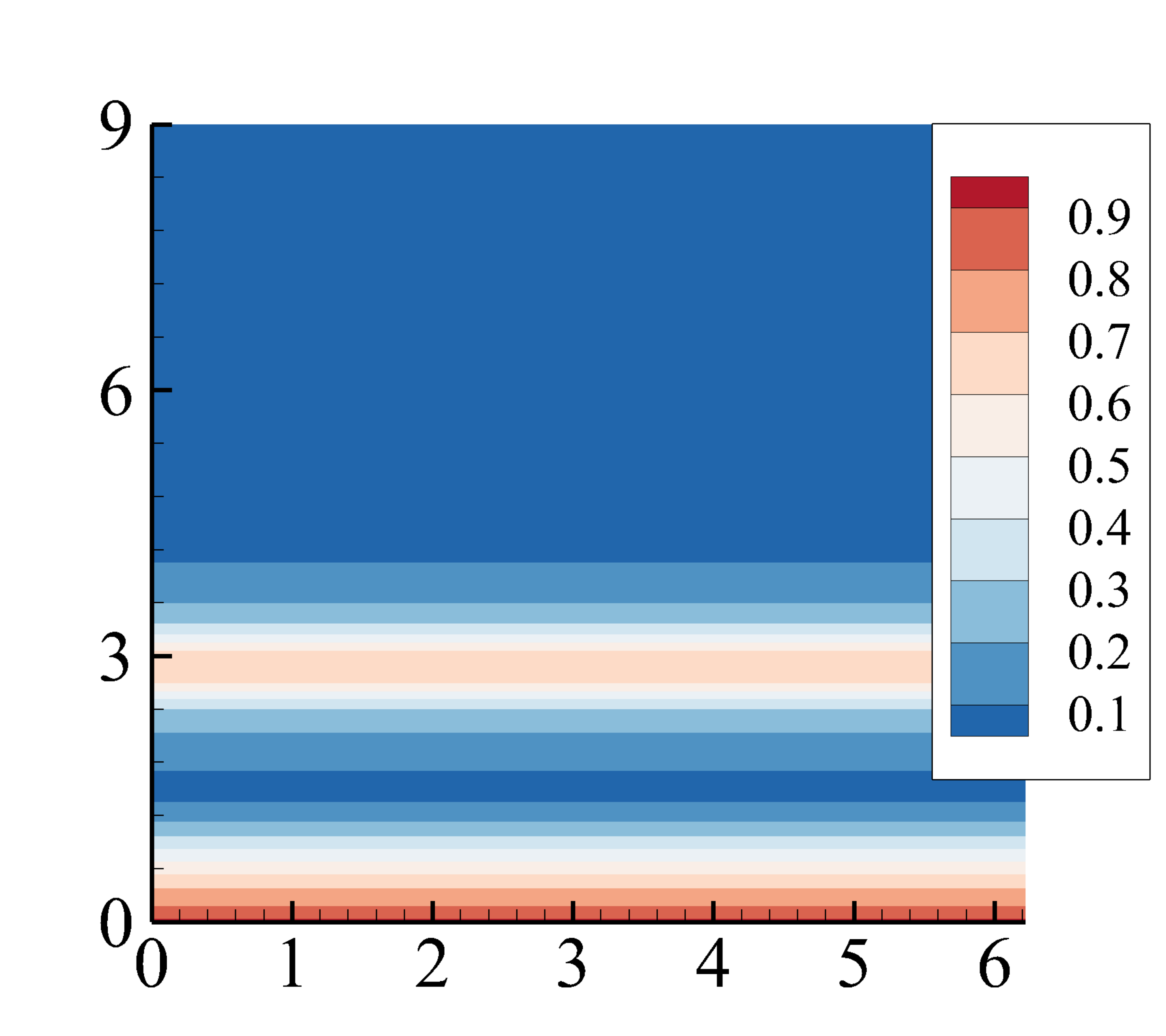}
\put(-80,85){2D Mack modes ($r_t=0$)}
\put(-93,76){\scriptsize{($a$)}}
\put(-93,40){\scriptsize $y$}
\put(-47,-5){\scriptsize $z$}
\put(-52,76){\color{black}\scriptsize NRM}
}
\hspace{-0.15cm}
\subfigure{
\includegraphics[width=0.24\textwidth,height=0.2\textwidth,trim=-0cm  0.2cm -0cm -0.1cm]{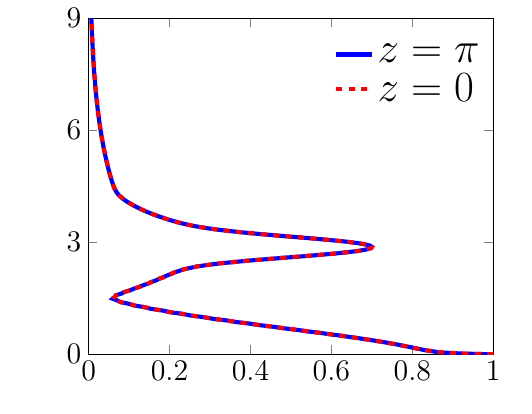}
\put(-90,76){\scriptsize{($b$)}}
\put(-90,40){\scriptsize $y$}
\put(-47,-5){\scriptsize $|p_s|$}
\put(-50,76){\color{black}\scriptsize NRM}
}
\hspace{0.1cm}
\subfigure{
\includegraphics[width=0.24\textwidth]{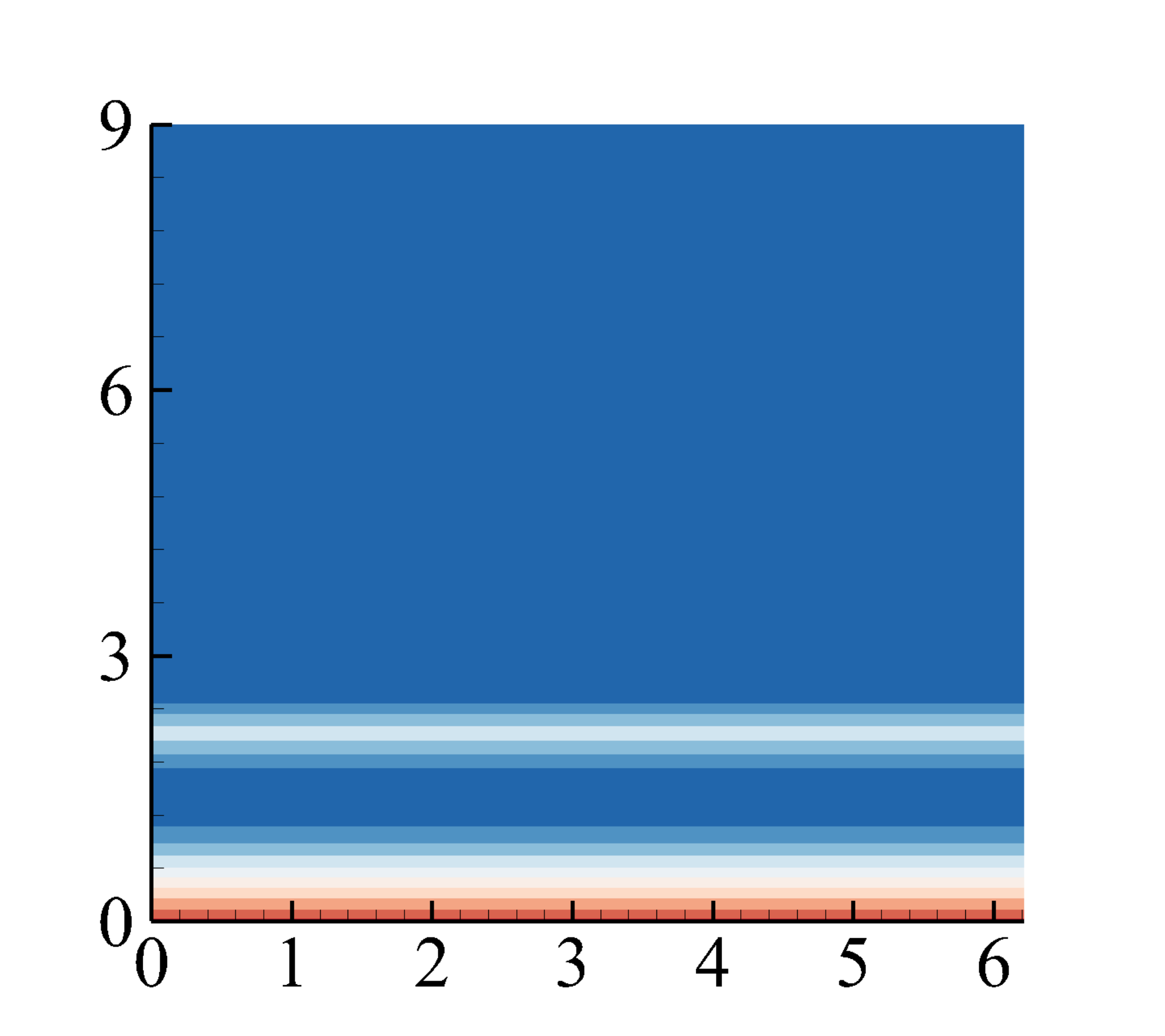}
\put(-93,76){\scriptsize{($c$)}}
\put(-93,40){\scriptsize $y$}
\put(-47,-5){\scriptsize $z$}
\put(-52,76){\color{black}\scriptsize RM}
}
\hspace{-0.15cm}
\subfigure{
\includegraphics[width=0.24\textwidth,height=0.2\textwidth,trim=-0cm  0.2cm -0cm -0.1cm]{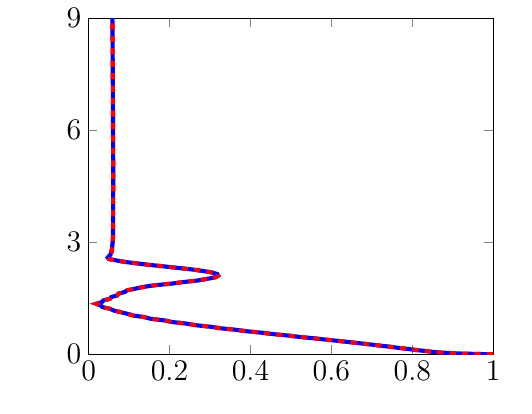}
\put(-90,76){\scriptsize{($d$)}}
\put(-90,40){\scriptsize $y$}
\put(-47,-5){\scriptsize $|p_s|$}
\put(-48,76){\color{black}\scriptsize RM}
}
\subfigure{
\includegraphics[width=0.24\textwidth]{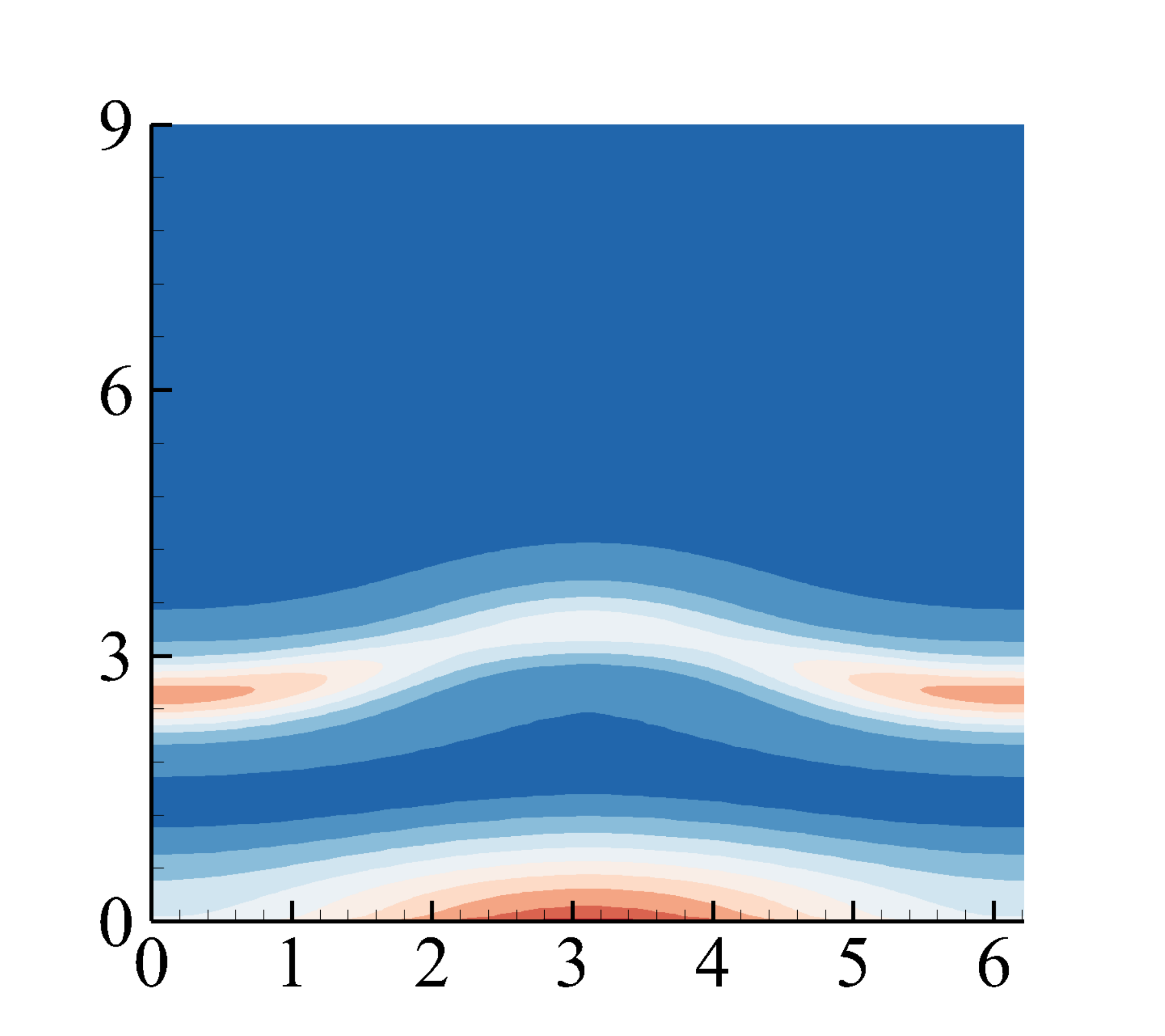}
\put(-80,85){Modified Mack modes ($r_t=1.1$)}
\put(-93,76){\scriptsize{($e$)}}
\put(-93,40){\scriptsize $y$}
\put(-47,-5){\scriptsize $z$}
\put(-52,76){\color{black}\scriptsize NRM}
}
\hspace{-0.15cm}
\subfigure{
\includegraphics[width=0.24\textwidth,height=0.2\textwidth,trim=-0cm  0.2cm -0cm -0.1cm]{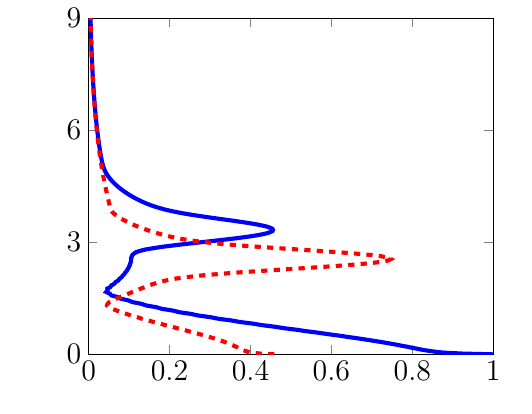}
\put(-50,76){\color{black}\scriptsize NRM}
\put(-90,76){\scriptsize{($f$)}}
\put(-90,40){\scriptsize $y$}
\put(-47,-5){\scriptsize $|p_s|$}
}
\hspace{0.1cm}
\subfigure{
\includegraphics[width=0.24\textwidth]{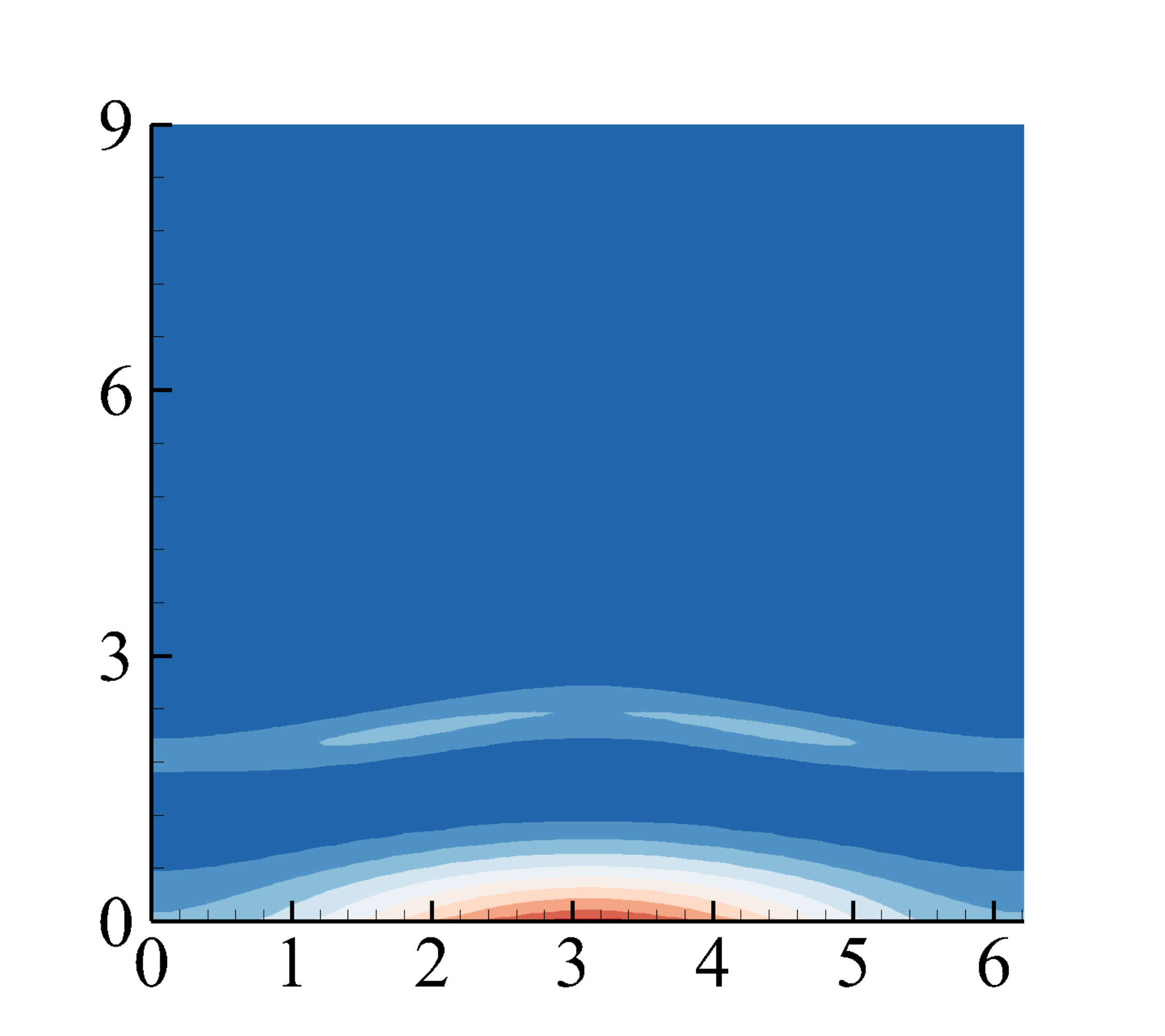}
\put(-93,76){\scriptsize{($g$)}}
\put(-93,40){\scriptsize $y$}
\put(-47,-5){\scriptsize $z$}
\put(-52,76){\color{black}\scriptsize RM}
}
\hspace{-0.15cm}
\subfigure{
\includegraphics[width=0.24\textwidth,height=0.2\textwidth,trim=-0cm  0.2cm -0cm -0.1cm]{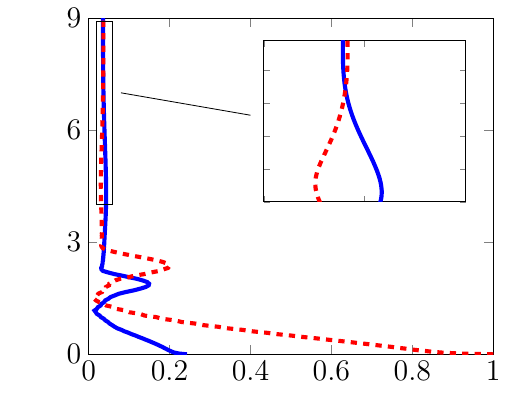}
\put(-90,76){\scriptsize{($h$)}}
\put(-90,40){\scriptsize $y$}
\put(-47,-5){\scriptsize $|p_s|$}
\put(-48,76){\color{black}\scriptsize RM}
}
\captionsetup{justification=raggedright}
\caption{($a$, $c$, $e$, $g$) Contours of $|p_s|$ and ($b$, $d$, $f$, $h$) local profiles at $z=0$ and $\pi.$ ($a$, $b$, $e$, $f$) non-radiating modes (NRM), $\alpha=0.59;$ ($c$, $d$, $g$, $h$) radiating modes (RM), $\alpha=0.9.$ The parameters are $R_\Lambda=10029,$ $\mathcal{M}_\infty=5$ and $T_w=0.17T_{ad}$.
}
\label{Mach-m}
\end{figure}

Figure \ref{Mach-m} displays the pressure eigenfunctions of the two-dimensional Mack modes ($r_t=0$) and of the modified Mack modes for $r_t=1.1$. The latter morph continuously from the former, and are therefore referred to as modified Mack modes rather than a distinct family of instability modes. In the absence of streaks, the conventional non-radiating Mack modes are completely trapped in the boundary layer (figure \ref{Mach-m}$a,b$). The radiating modes also reside primarily in the boundary layer, but their pressure propagates to the far field (figure \ref{Mach-m}$c,d$). In the presence of streaks, the modes become fully three-dimensional: the eigenfunctions of the non-radiating modes (figure \ref{Mach-m}$e,f$) and radiating modes (figure \ref{Mach-m}$g,h$) are modulated along the spanwise direction within the boundary layer. Outside of the boundary layer, the pressure of the radiating mode does not vary along the spanwise direction as only the two-dimensional component emits a sound wave.

\begin{figure}
\centering
\subfigure{
\includegraphics[width=0.42\textwidth]{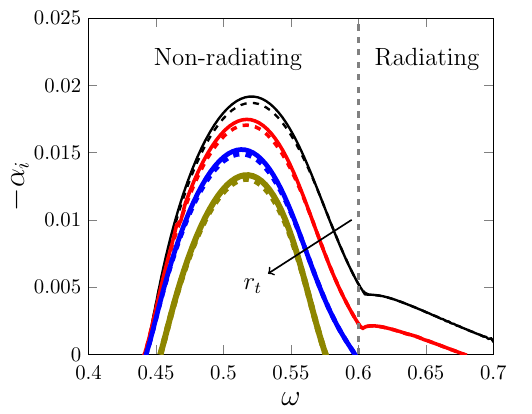}
\put(-165,118){($a$)}
}
\subfigure{
\includegraphics[width=0.42\textwidth]{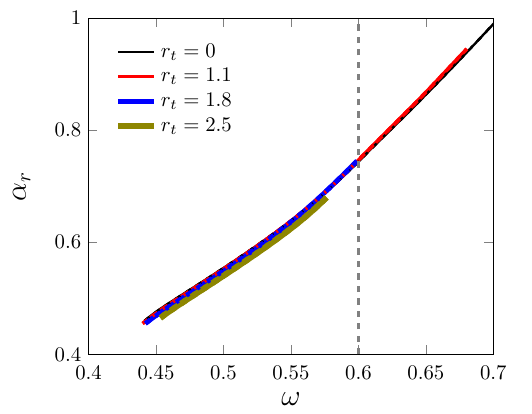}
\put(-165,118){($b$)}
}
\captionsetup{justification=raggedright}
\caption{Effect of the amplitude of high-supersonic streaks on Mack modes. ($a$) Spatial growth rate $-\alpha_i$; ($b$) spatial wavenumber $\alpha_r$. The dashed and solid lines denote the Gaster transformation \eqref{eq:first_order_secondary} and XLZW transformation \eqref{eq:second_order_secondary}, respectively. The vertical grey dashed lines correspond to the critical phase velocity, $c_{r,c} = 1 - 1/\mathcal{M}_\infty$, and separate the non-radiating and radiating modes. The  parameters are $R_\Lambda=10029,$ $\mathcal{M}_\infty=5$ and $T_w=0.17T_{ad}.$}
\label{fig:2D-mod}
\end{figure}
Figure \ref{fig:2D-mod} presents the spatial growth rate $-\alpha_i$ and the spatial wavenumber $\alpha_r$ of the modified Mack modes obtained using the Gaster transformation \eqref{eq:first_order_secondary} and the XLZW transformation \eqref{eq:second_order_secondary}. For the cases considered, the Gaster transformation is sufficiently accurate.  The spatial growth rate decreases as the streak amplitude increases. The maximum spatial growth rate occurs at a frequency $\omega = 0.52$ for the non-radiating Mack mode. At this frequency, the real part of the wavenumber is 0.6, indicating that the dominant streamwise wavelength is 1.7 times the spanwise wavelength of the streaks.

\subsection{Temporal and spatial secondary-instability modes}
\label{sec:second-family}

The secondary instability of the FVD-induced streaks in high-supersonic boundary layers over cooled walls is now reported for the first time. While the modified radiating Mack modes are completely stabilised by the streaks, a new family of three-dimensional secondary-instability modes comes into existence when the FVD becomes sufficiently intense, specifically for $r_t>1.2$. This $r_t$ value corresponds to a turbulence level $Tu=0.03$\%, which is representative of typical quiet flight conditions. Below this threshold, no secondary-instability modes are detected. The secondary-instability modes are named peak or valley modes according to the primary location of their eigenfunctions in the cross-sections of the streaks, as indicated in figure \ref{fig:utrms-tw}$(e)$.
A mode is further classified as sinuous if its spanwise velocity obeys
$\hat w_s(y,z)=\hat w_s(y,2\pi-z)$ and varicose if its spanwise velocity obeys
$\hat w_s(y,z)=-\hat w_s(y,2\pi-z)$. Equivalently, the pressure of a sinuous mode satisfies $\hat p_s(y,z)=-\hat p_s(y,2\pi-z)$ and the pressure of a varicose mode satisfies $\hat p_s(y,z)=\hat p_s(y,2\pi-z)$ since the spanwise velocity $\hat w_s$ and the pressure $\hat p_s$ have opposite symmetry properties according to the momentum equations.

\begin{figure}
\centering
\subfigure{
\includegraphics[width=0.315\textwidth]{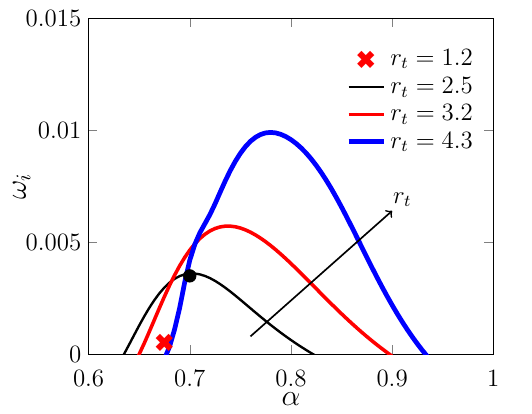}
\put(-120,95){($a$)}
\put(-89,95){ \scriptsize Valley varicose modes}
\put(-55,1){\makebox(0,0){\colorbox{white}{\footnotesize $\alpha$}}}
\put(-118,55){\makebox(0,0){\rotatebox{90}{\colorbox{white}{\footnotesize $\omega_i$}}}}
}
\subfigure{
\includegraphics[width=0.315\textwidth]{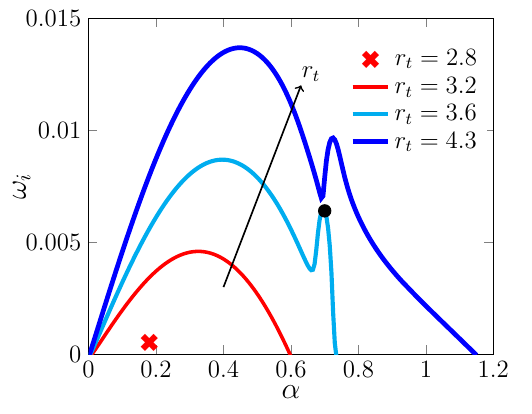}
\put(-120,95){($b$)}
\put(-85,95){ \scriptsize Peak sinuous modes}
\put(-55,3){\makebox(0,0){\colorbox{white}{\footnotesize $\alpha$}}}
\put(-118,55){\makebox(0,0){\rotatebox{90}{\colorbox{white}{\footnotesize $\omega_i$}}}} }
\subfigure{
\includegraphics[width=0.315\textwidth]{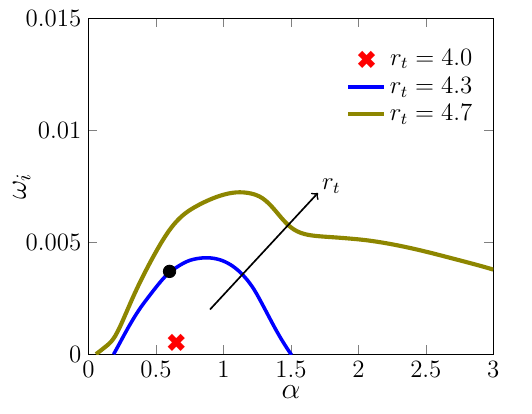}
\put(-120,95){($c$)}
\put(-87,95){ \scriptsize Peak varicose modes}
\put(-55,3){\makebox(0,0){\colorbox{white}{\footnotesize $\alpha$}}}
\put(-118,55){\makebox(0,0){\rotatebox{90}{\colorbox{white}{\footnotesize $\omega_i$}}}} }

\subfigure{
\includegraphics[width=0.315\textwidth]{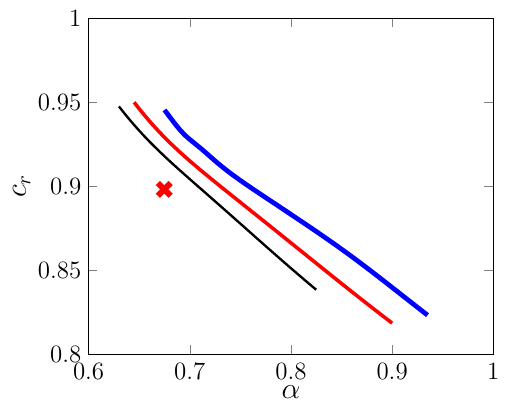}
\put(-120,95){($d$)}
\put(-55,2){\makebox(0,0){\colorbox{white}{\footnotesize $\alpha$}}}
\put(-118,50){\makebox(0,0){\rotatebox{90}{\colorbox{white}{\footnotesize $c_r$}}}}
}
\subfigure{
\includegraphics[width=0.315\textwidth]{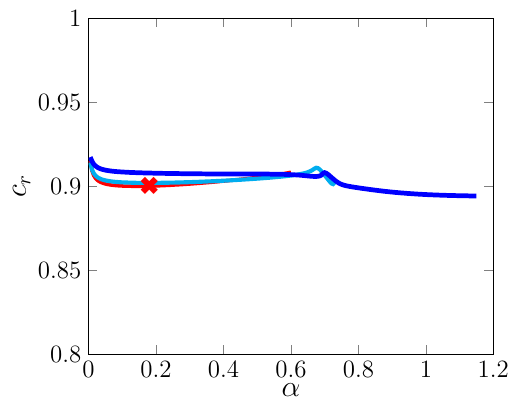}
\put(-120,95){($e$)}
\put(-55,3){\makebox(0,0){\colorbox{white}{\footnotesize $\alpha$}}}
\put(-118,50){\makebox(0,0){\rotatebox{90}{\colorbox{white}{\footnotesize $c_r$}}}} }
\subfigure{
\includegraphics[width=0.315\textwidth]{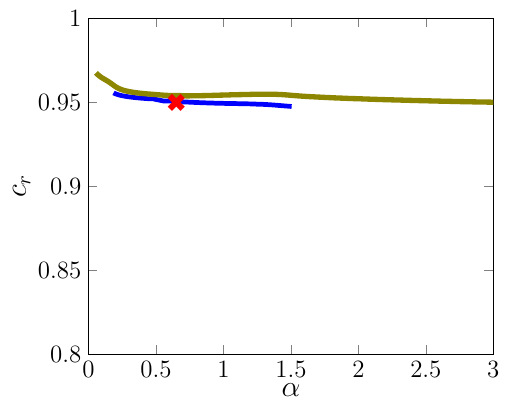}
\put(-120,95){($f$)}
\put(-55,3){\makebox(0,0){\colorbox{white}{\footnotesize $\alpha$}}}
\put(-118,50){\makebox(0,0){\rotatebox{90}{\colorbox{white}{\footnotesize $c_r$}}}} }

\hspace{0.3cm}
\subfigure{
\includegraphics[width=0.31\textwidth]{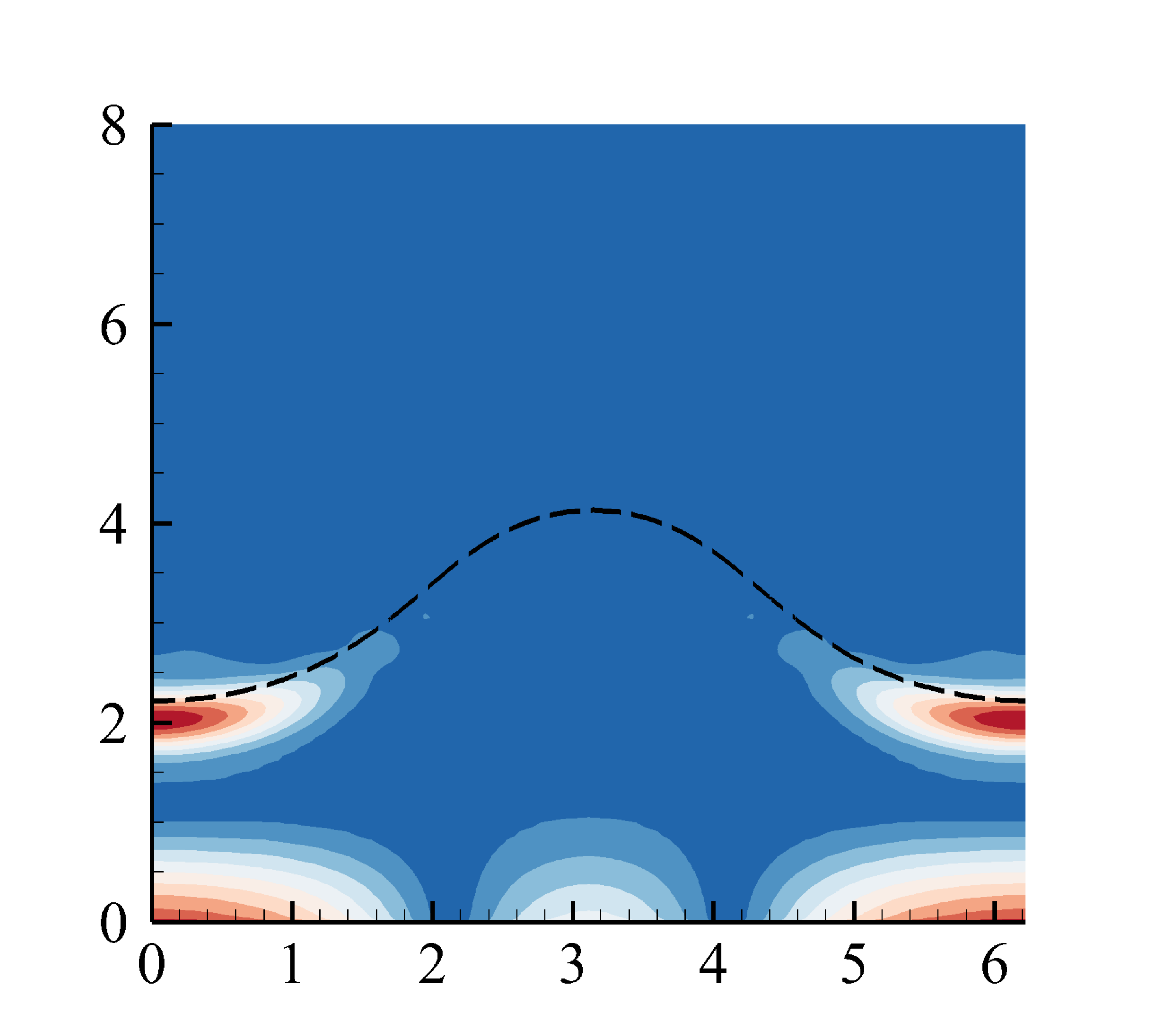}
\put(-123,95){{($g$)}}
\put(-120,55){\footnotesize $y$}
\put(-60,-3){\footnotesize $z$}
}\hspace{0.1cm}
\subfigure{
\includegraphics[width=0.31\textwidth]{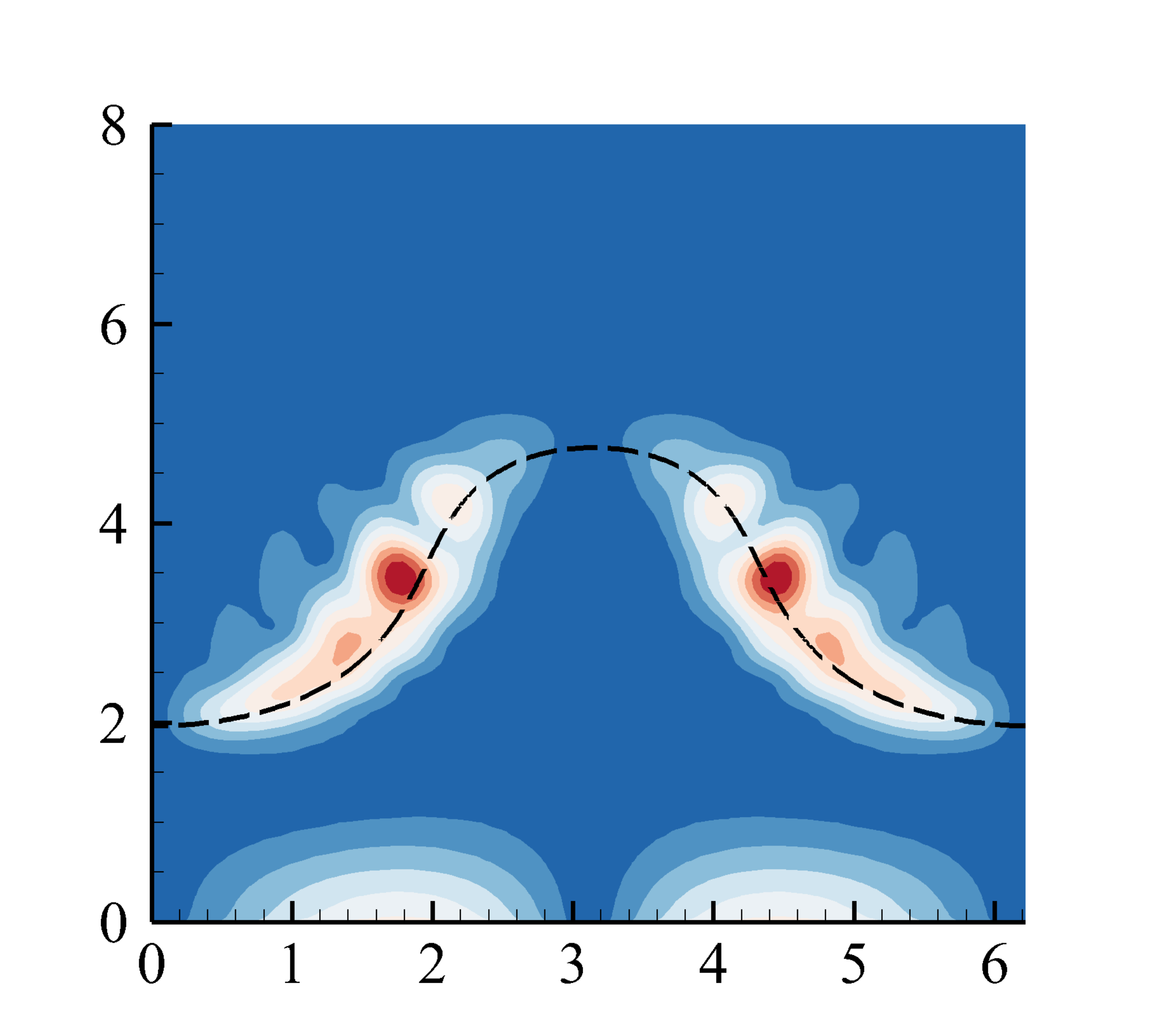}
\put(-123,95){{($h$)}}
\put(-120,55){\footnotesize $y$}
\put(-60,-3){\footnotesize $z$}
}\hspace{-0.1cm}
\subfigure{
\includegraphics[width=0.31\textwidth]{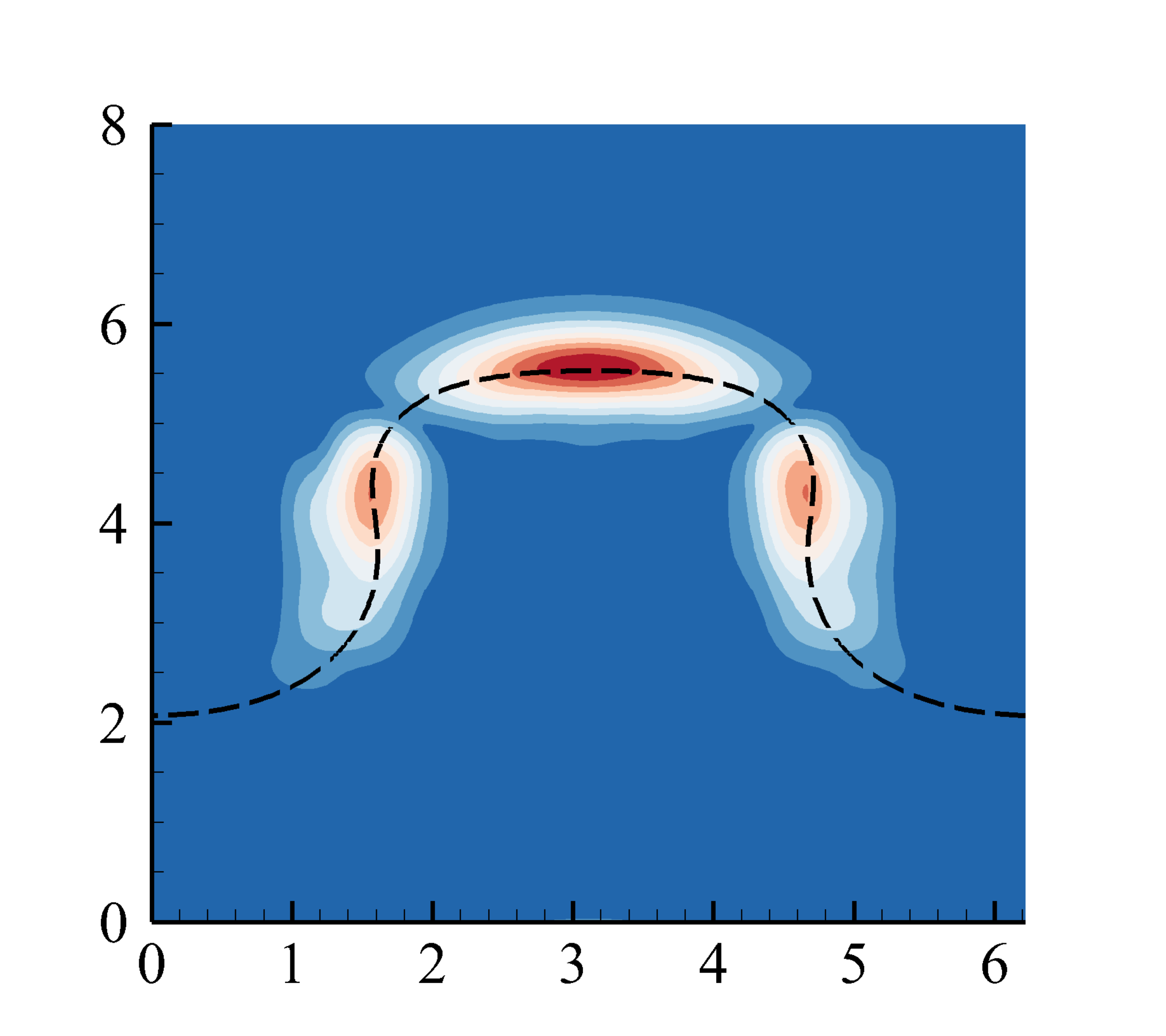}
\put(-123,95){{($i$)}}
\put(-120,55){\footnotesize $y$}
\put(-60,-3){\footnotesize $z$}
}

\hspace{0.3cm}
\subfigure{
\includegraphics[width=0.31\textwidth]{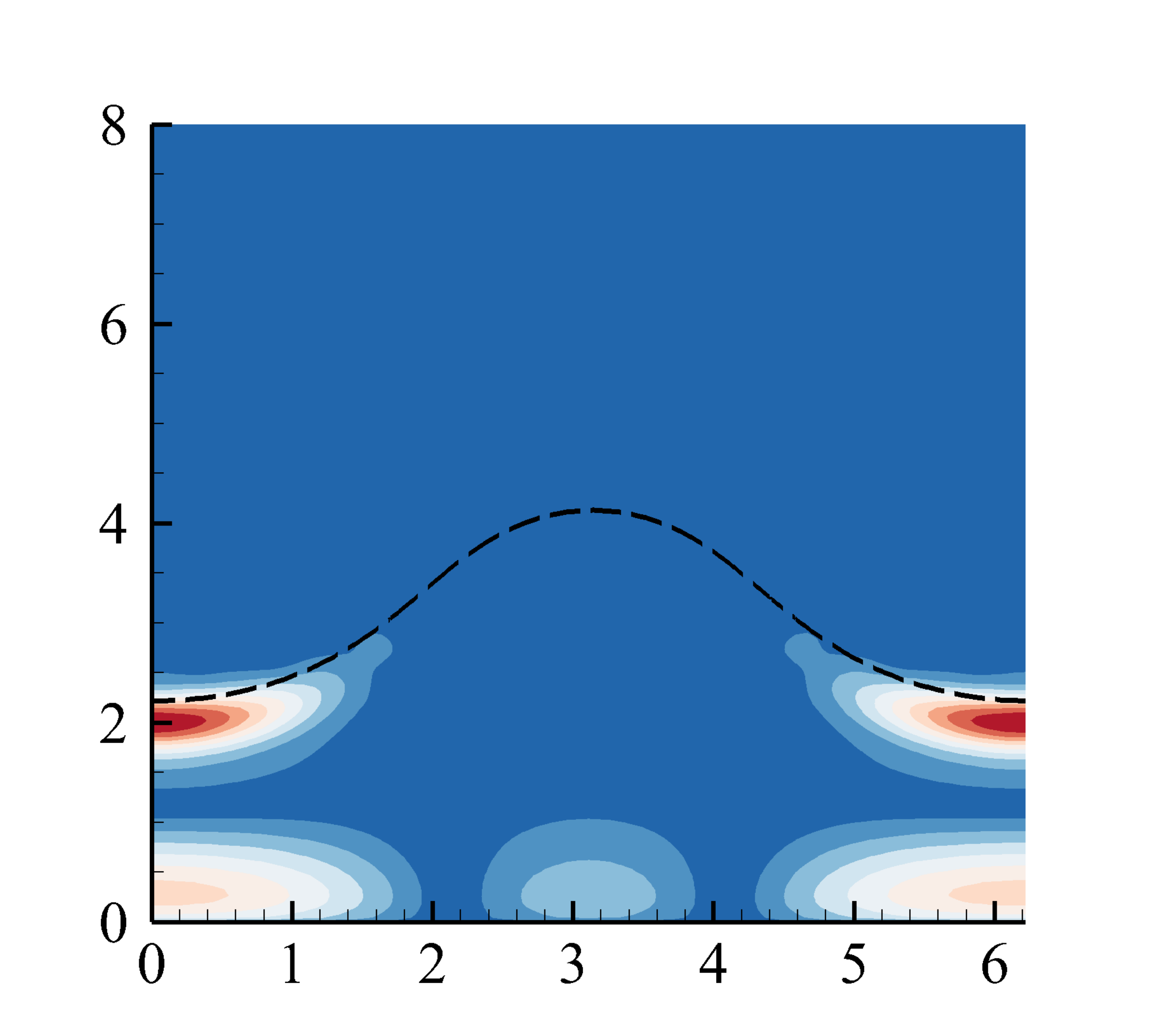}
\put(-123,95){{($j$)}}
\put(-120,55){\scriptsize $y$}
\put(-60,-3){\scriptsize $z$}
}\hspace{0.1cm}
\subfigure{
\includegraphics[width=0.31\textwidth]{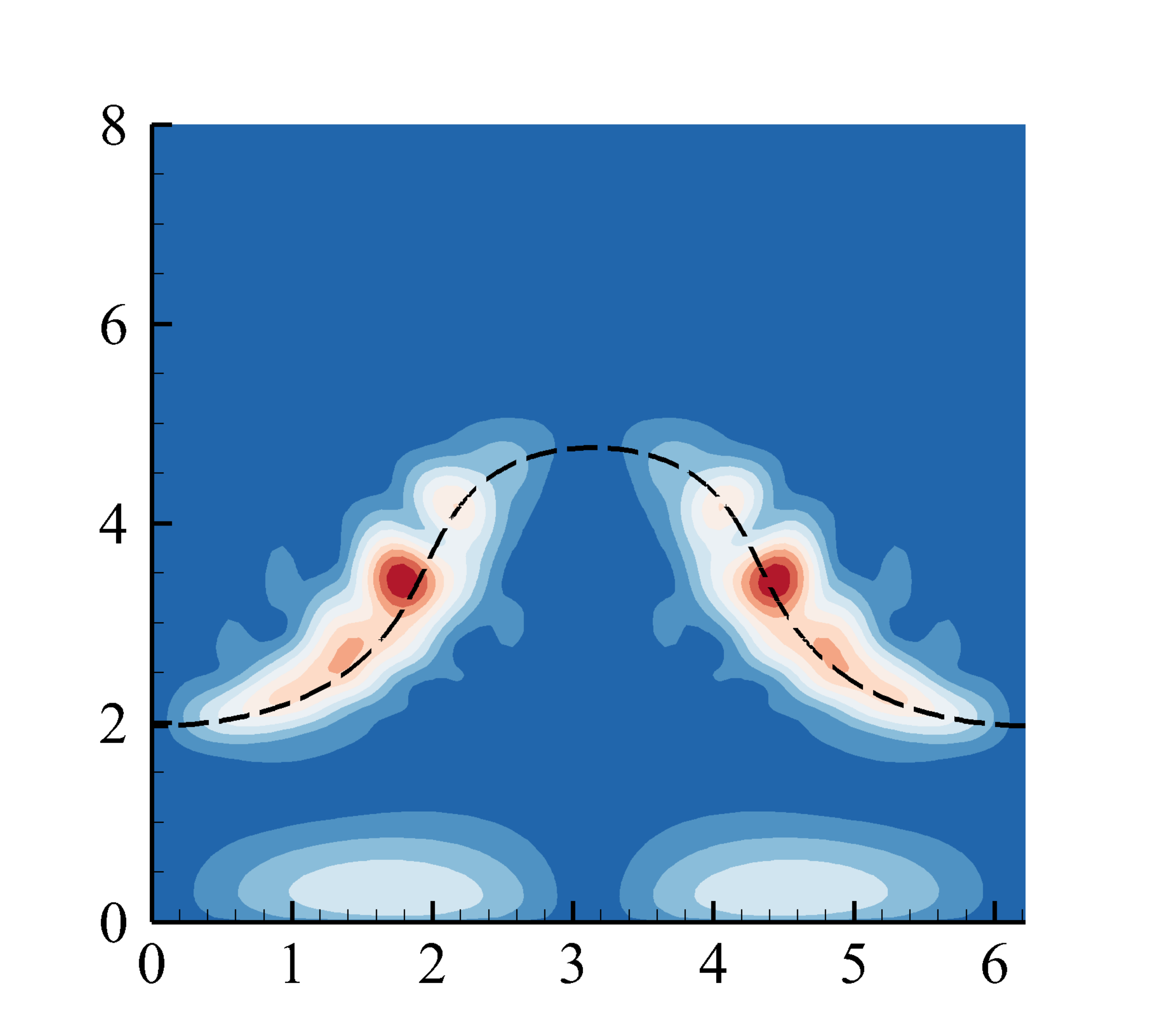}
\put(-123,95){{($k$)}}
\put(-120,55){\scriptsize $y$}
\put(-60,-3){\scriptsize $z$}
}\hspace{-0.1cm}
\subfigure{
\includegraphics[width=0.31\textwidth]{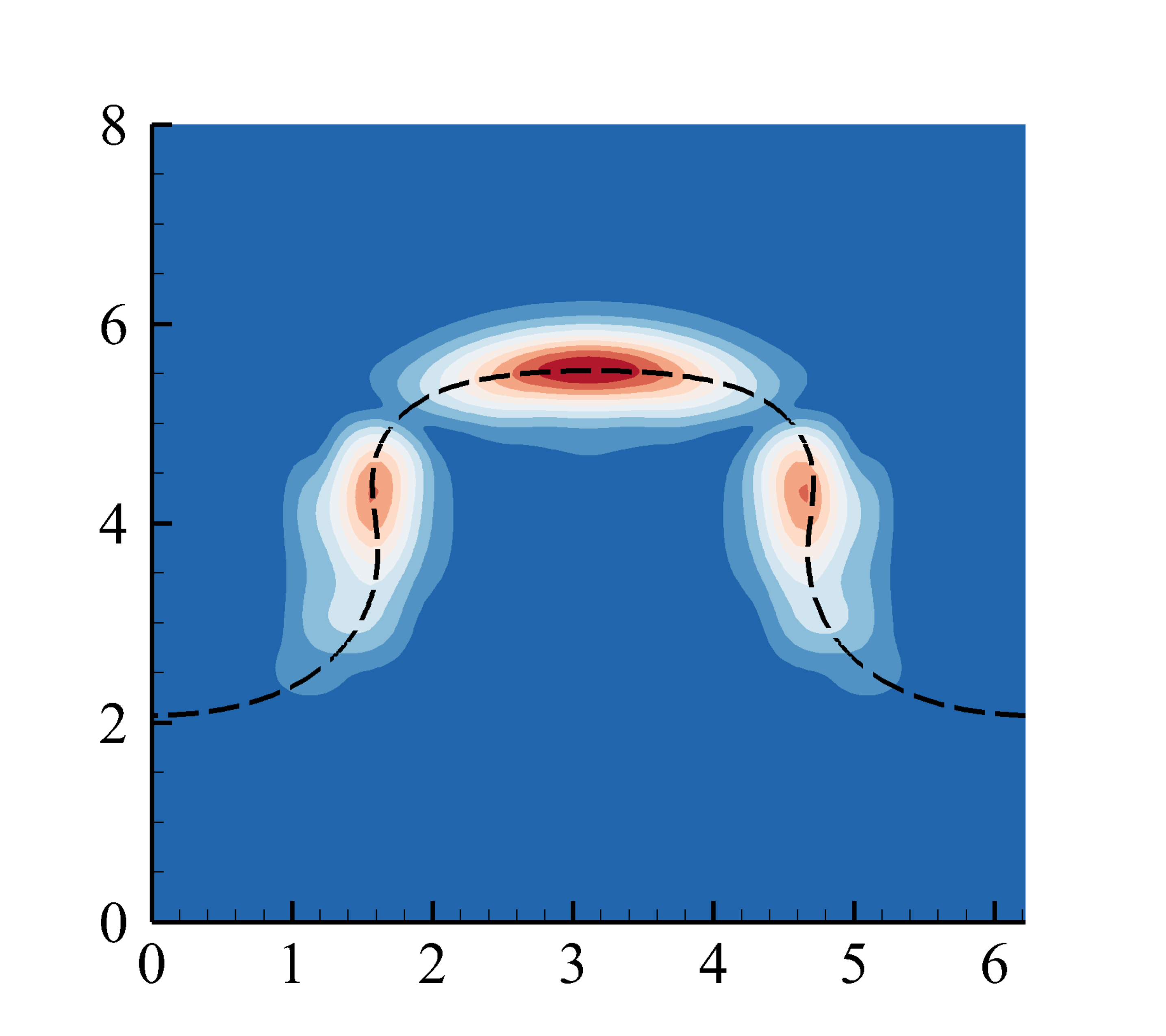}
\put(-123,95){{($l$)}}
\put(-120,55){\scriptsize $y$}
\put(-60,-3){\scriptsize $z$}
}

\captionsetup{justification=raggedright}
\caption{Secondary instability of high-supersonic streaks: ($a$-$c$) Temporal growth rate; ($d$-$f$) phase velocity. Eigenfunctions of the secondary-instability modes (normalised pressure ($g$-$i$) and temperature ($j$-$l$), where the same colour bar as in figure \ref{Mach-m} is used): ($g,j$) $\alpha=0.7,$ $r_t=2.5;$ ($h,k$) $\alpha=0.7,$ $r_t=3.6;$ ($i,l$) $\alpha=0.6,$ $r_t=4.3$. The dashed lines in ($g$-$l$) mark the critical layers where $\tilde u=c_{r,c}.$ The dots in panels $(a$-$c)$ denote the cases shown in panels $(g$-$l)$ and the crosses mark the critical $r_t$ at which $\omega_i=5\times10^{-4}.$ The parameters are $R_\Lambda=10029,$ $\mathcal{M}_\infty=5$ and $T_w=0.17T_{ad}.$}
\label{sec-sup}
\end{figure}
Figure \ref{sec-sup} shows the growth rates and phase velocities of the secondary-instability modes. The onset of secondary instability is identified by using the small threshold $\omega_i = 5 \times 10^{-4}$, which distinguishes growth from neutral stability. The first secondary mode to emerge for $r_t=1.2$ is a newly identified valley varicose mode (figure \ref{sec-sup}$a$). For $r_t=2.5$, i.e. $Tu=0.07\%$, the growth rate of the valley mode becomes comparable to those of the modified Mack modes. For the cases studied, this valley mode is only unstable in a certain wavenumber range, i.e. $0.6<\alpha<1.0$. It is non-radiating as its phase velocity is larger than the critical phase velocity $c_{r,c}=1-1/\mathcal{M}_\infty=0.8$ for $\mathcal{M}_\infty=5$ (figure \ref{sec-sup}$d$). Increasing the FVD level leads to larger growth rates and shifts the instability range to larger wavenumbers. The dominance of the valley mode is a robust physical feature, insensitive to the specific cutoff criterion. The significance of this mode is that it is the first secondary-instability branch to become unstable as the FVD intensity $r_t$ is increased, while the peak sinuous and varicose modes remain stable. Since this ordering of the secondary modes is established here for strongly distorted streaks and for a realistic FVD configuration, the emergence of the valley mode is likely to impact the transition route by which the FVD-induced streaks may undergo breakdown to turbulence. The decay of the valley varicose mode at smaller $r_t$ could be a topic worth further investigation, as discussed by \cite{caillaud2025}.

The peak sinuous and varicose modes become unstable when  $r_t=2.8$ and $r_t=4.0,$ respectively. Similar modes were reported by \citet{paredes2019instability} for streaks developed from optimal disturbances. Thanks to our receptivity--instability framework, we are able to relate these high-supersonic modes to the intensity of physically-realisable FVD. As shown in figure \ref{sec-sup}($b$), the peak sinuous mode is the most unstable among the secondary-instability modes for the same, sufficiently large, $r_t$. Its growth rate curves have two maxima for high levels of FVD, while only one maximum appeared in the curves computed by \citet{paredes2019instability}. Figure \ref{sec-sup}($c$,$f$) shows the growth rate and phase velocity of the peak varicose mode, which appears when $r_t > 4.0$. This mode is less unstable than the peak sinuous mode and its wavenumber range of instability expands substantially as $r_t$ increases. For \( r_t > 4.7 \), further instability branches emerge (not shown), but they have not been observed in either experimental or DNS studies. The linear analysis may be invalid for these modes because the distortion of the streaks caused by the growing dominant modes may be too large. Because the eigenfunction of the dominant mode is not confined to the peak but extends across the flanks of the streaks towards the valley (figure \ref{sec-sup}$h,k$), sufficiently large growth of this mode may distort the streaks enough to affect the valley region and induce the breakdown to turbulence there. Further investigation is required to determine how such distortion affects the valley modes, how these modes interact nonlinearly and how the instability waves feedback onto the streaks \citep{brandt2002transition,hack2014streak}.
The phase speed of the peak sinuous mode is about 0.9, while that of the peak varicose mode is about 0.95, i.e. the largest among the three secondary modes (figure \ref{sec-sup}$e$-$f$). The peak sinuous and varicose modes display a very mild dispersive behaviour as their phase speeds are weakly dependent on the wavenumber, unlike the valley varicose mode whose phase velocity decreases significantly with the wavenumber (figure \ref{sec-sup}$d$).

The eigenfunctions of the secondary-instability modes are displayed in figure \ref{sec-sup}($g$-$i$) (pressure) and in figure \ref{sec-sup}($j$-$l$) (temperature). Unlike the incompressible modes reported in \citet{andersson2001breakdown}, whose eigenfunctions concentrate in the streak peak region, the valley varicose mode resides primarily near the wall and in thin regions near the critical layer ($\tilde u=c_{r,c}$). A notable feature of this mode is the presence of high-temperature fluctuations near the wall. The peak sinuous and varicose modes are akin to those reported in  \cite{andersson2001breakdown} and \cite{paredes2019instability}, in that they concentrate in the low-speed region and in the bulk of the boundary layer (figure \ref{sec-sup}$h$-$i$). Different from the results of \cite{andersson2001breakdown} and \cite{paredes2019instability}, the eigenfunctions of the present peak sinuous modes also extend into the near-wall layer (figure \ref{sec-sup}$h$-$k$). Our study is the first to have identified these three families of secondary-instability modes in an FVD-induced high-supersonic streaky boundary layer.
\begin{figure}
\centering
\subfigure{
\includegraphics[width=0.315\textwidth]{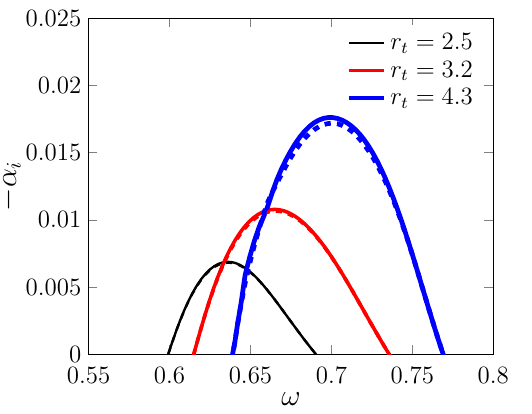}
\put(-120,95){($a$)}
\put(-85,95){ \scriptsize Valley varicose modes}
\put(-55,1){\makebox(0,0){\colorbox{white}{\footnotesize $\omega$}}}
\put(-121,52){\makebox(0,0){\rotatebox{90}{\colorbox{white}{\footnotesize $-\alpha_i$}}}}
}
\subfigure{
\includegraphics[width=0.315\textwidth]{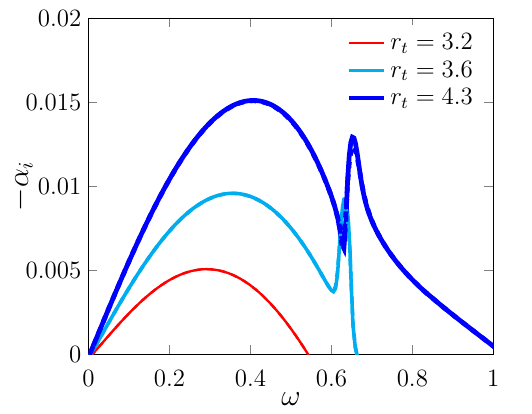}
\put(-120,95){($b$)}
\put(-85,95){ \scriptsize Peak sinuous modes}
\put(-55,1){\makebox(0,0){\colorbox{white}{\footnotesize $\omega$}}}
\put(-119,52){\makebox(0,0){\rotatebox{90}{\colorbox{white}{\footnotesize $-\alpha_i$}}}}
}
\subfigure{
\includegraphics[width=0.315\textwidth]{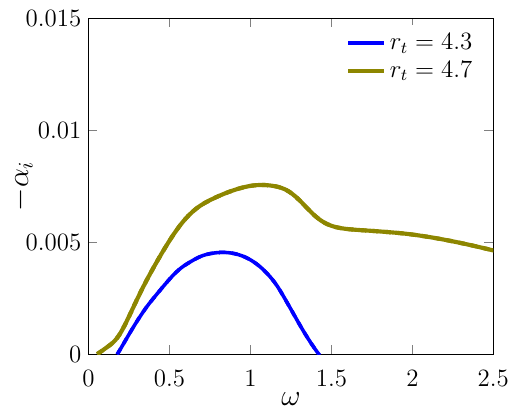}
\put(-120,95){($c$)}
\put(-85,95){ \scriptsize Peak varicose modes}
\put(-55,1){\makebox(0,0){\colorbox{white}{\footnotesize $\omega$}}}
\put(-119,52){\makebox(0,0){\rotatebox{90}{\colorbox{white}{\footnotesize $-\alpha_i$}}}}
}
\subfigure{
\includegraphics[width=0.315\textwidth]{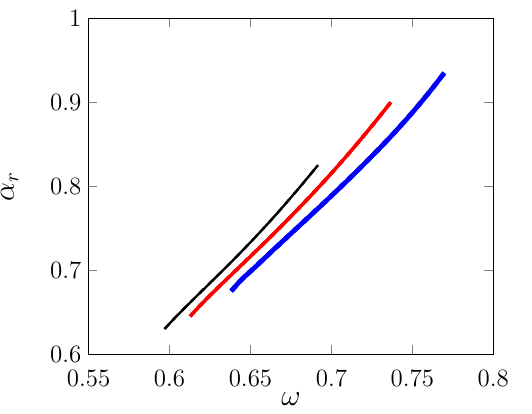}
\put(-120,95){($d$)}
\put(-55,1){\makebox(0,0){\colorbox{white}{\footnotesize $\omega$}}}
\put(-119,52){\makebox(0,0){\rotatebox{90}{\colorbox{white}{\footnotesize $\alpha_r$}}}}
}
\subfigure{
\includegraphics[width=0.315\textwidth]{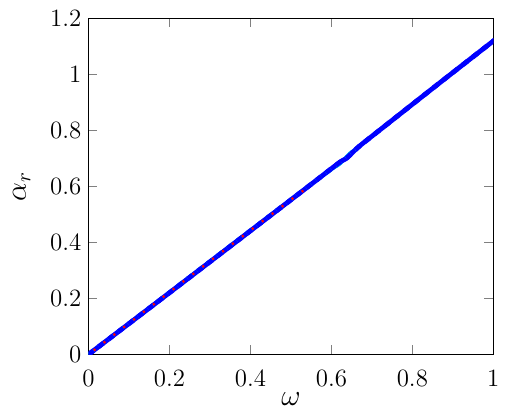}
\put(-120,95){($e$)}
\put(-55,1){\makebox(0,0){\colorbox{white}{\footnotesize $\omega$}}}
\put(-119,52){\makebox(0,0){\rotatebox{90}{\colorbox{white}{\footnotesize $\alpha_r$}}}}
}
\subfigure{
\includegraphics[width=0.315\textwidth]{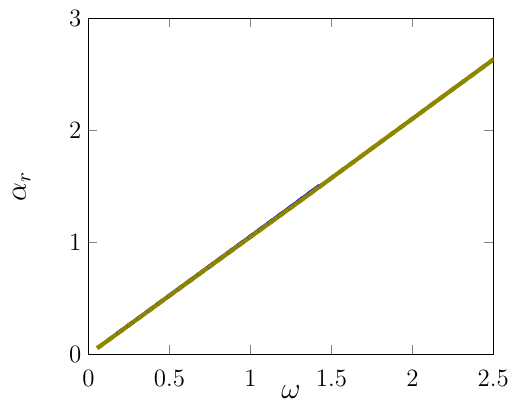}
\put(-120,95){($f$)}
\put(-55,1){\makebox(0,0){\colorbox{white}{\footnotesize $\omega$}}}
\put(-119,52){\makebox(0,0){\rotatebox{90}{\colorbox{white}{\footnotesize $\alpha_r$}}}}
}
\captionsetup{justification=raggedright}
\caption{Secondary instability of high-supersonic streaks: ($a$-$c$) Spatial growth rate $-\alpha_i$; ($d$-$f$) spatial wavenumber $\alpha_r.$ The dashed and solid lines denote the Gaster transformation \eqref{eq:first_order_secondary} and XLZW transformation \eqref{eq:second_order_secondary}, respectively. The parameters are $R_\Lambda=10029,$ $\mathcal{M}_\infty=5$ and $T_w=0.17T_{ad}.$}
\label{sec-spa}
\end{figure}

Figure \ref{sec-spa} illustrates the results of the spatial secondary instability obtained from the temporal stability calculations by using the Gaster transformation \eqref{eq:first_order_secondary} and the XLZW transformation \eqref{eq:second_order_secondary}. Figure \ref{sec-spa}($a$-$c$) presents the spatial growth rates of the valley and peak modes. As the FVD intensity $r_t$ varies, both the maximum growth rate and its corresponding frequency increase.
Figure \ref{sec-spa}($d$-$f$) also shows the wavenumber $\alpha_r$ for both mode types. Again, the XLZW and Gaster transformations yield nearly identical results. For the valley modes, the frequency ranges from 0.55 to 0.8. For the peak modes, it spans from 0 to 1 in the sinuous branch (three-line overlaps) and it extends beyond 2.5 in the varicose branch (two-line overlaps).

\begin{figure}
\centering
\subfigure{
\includegraphics[width=0.42\textwidth]{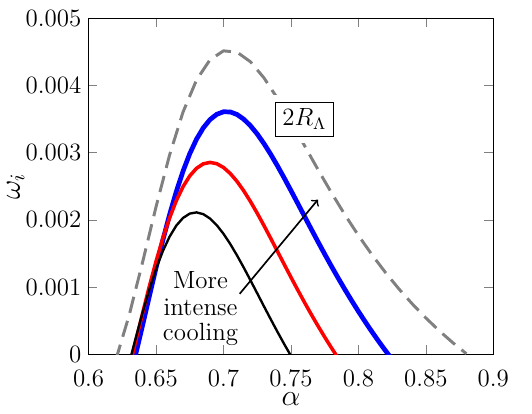}
\put(-165,118){($a$)}
\put(-71,3){\makebox(0,0){\colorbox{white}{\footnotesize $\alpha$}}}
\put(-158,70){\makebox(0,0){\rotatebox{90}{\colorbox{white}{\footnotesize $\omega_i$}}}}
}
\subfigure{
\includegraphics[width=0.42\textwidth]{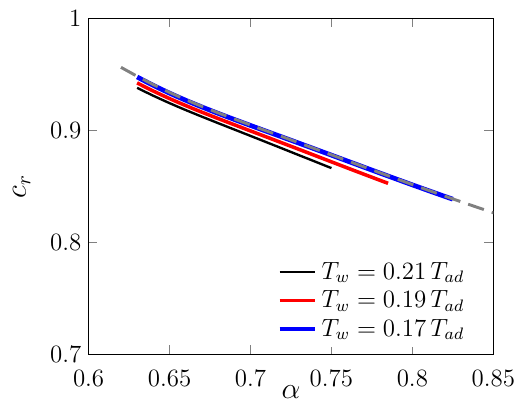}
\put(-165,118){($b$)}
\put(-71,3.5){\makebox(0,0){\colorbox{white}{\footnotesize $\alpha$}}}
\put(-158,70){\makebox(0,0){\rotatebox{90}{\colorbox{white}{\footnotesize $c_r$}}}}
}

\subfigure{
\includegraphics[width=0.42\textwidth]{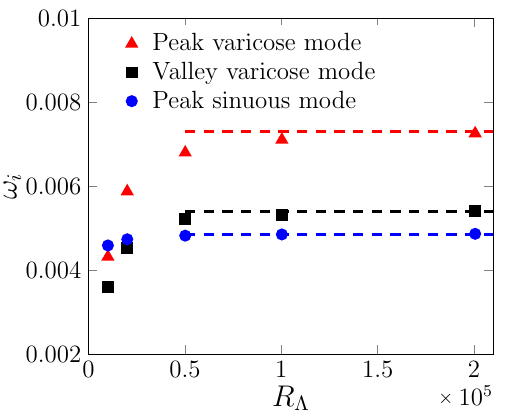}
\put(-165,118){($c$)}
\put(-71,3){\makebox(0,0){\colorbox{white}{\footnotesize $R_\Lambda$}}}
\put(-160,70){\makebox(0,0){\rotatebox{90}{\colorbox{white}{\footnotesize $\omega_i$}}}}
}
\captionsetup{justification=raggedright}
\caption{Wall-temperature effect on ($a$) the temporal growth rate and ($b$) the phase velocity of the valley varicose mode. The parameters are $\mathcal{M}_\infty=5$ and $r_t=2.5.$ The solid lines are for $R_\Lambda=10029$. The dashed grey lines correspond to a Reynolds number $2R_{\Lambda}$ and $T_w=0.17T_{ad}.$ ($c$) Dependence of the growth rate on the Reynolds number $R_\Lambda$ at $T_w=0.17T_{ad}$. The turbulent intensities for valley varicose, peak sinuous and peak varicose modes are $r_t=2.5,$ 3.2 and 4.3, respectively. The dashed lines indicate the inviscid limits of the corresponding modes.
}
\label{fig:utrms-tweffect}
\end{figure}
Figure \ref{fig:utrms-tweffect}$(a,b)$ shows the wall-cooling effect on the growth rates and phase velocities of the valley varicose modes. These modes become more amplified and their phase velocity increases moderately as the wall temperature is reduced. This response to wall cooling is different from that of the Mack second modes. These valley modes are observed only under extreme wall-cooling conditions and are absent for $T_w/T_{ad}>0.3$. This behaviour is distinct from the second-mode frequency splitting reported in \citet{caillaud2025}, since the latter can exist even under adiabatic-wall conditions.

Figure \ref{fig:utrms-tweffect}($a,b$) further suggests that this family of instability modes is inviscid in character because, as the Reynolds number \(R_{\Lambda}\) is doubled, its growth rate increases, while the unstable band of phase velocities covers a wider range. Here we have increased \(R_{\Lambda}\) while keeping the Mach number and all the other non-dimensional parameters constant, as in \citet{viaro2019compressible}. The inviscid character of all three modes is confirmed by their growth rates and phase velocities approaching their inviscid limits when $R_\Lambda$ becomes sufficiently large. These limits are reached when $R_\Lambda$ is about $5\times 10^4$ for the peak sinuous mode, $10^5$ for the valley varicose mode, and $2\times 10^5$ for the peak varicose mode, as shown in figure \ref{fig:utrms-tweffect}($c$). Although the limiting growth rates at large Reynolds number are different and the modes differ in their spatial structures, localisation and growth characteristics, these differences alone do not imply distinct underlying physical mechanisms. For moderate \(R_{\Lambda}\), viscous effects are substantial and the wall-layer perturbations are fairly intense in contrast to the peak modes, which remain trapped in the core of the boundary layer (figure \ref{sec-sup}$h$,$i$). This feature of the valley mode may have significant implications for thermal protection.

\begin{figure}
\centering
\includegraphics[width=0.9\textwidth]{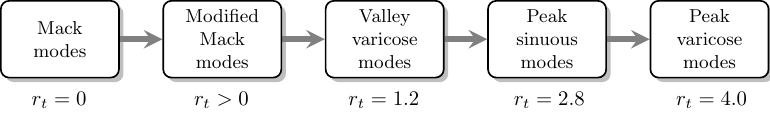}
\captionsetup{justification=raggedright}
\caption{Instability modes in unperturbed (\(r_t=0\)) and streaky cooled-wall high-supersonic boundary layers. The reported $r_t$ values   represent the threshold for the onset of the corresponding modes and the arrows indicate the direction of increasing $r_t$. The parameters are $R_\Lambda=10029$, $\mathcal{M}_\infty=5$ and $T_w=0.17T_{ad}.$}
\label{fig:sum}
\end{figure}
Figure \ref{fig:sum} summarises the sequence in which the modes come into existence in cooled-wall conditions as the FVD level is increased. This sequence pertains to the present FVD configuration and to streamwise locations and phases at which the streaks are highly distorted. The modified Mack second modes, the newly identified valley varicose modes, the peak sinuous modes and the peak varicose modes all reach comparable growth rates, provided that the FVD level is sufficiently high. The effects of variations in the streak phase, streamwise location and FVD configuration on the onset and relative importance of the different branches can be investigated within the present receptivity--instability framework, as performed in \cite{xu_zhang_wu_2017,xu2020gortler}. This future work can reveal whether the newly identified valley varicose mode stays dominant as the branch that emerges at the lowest $r_t$ across a broader range of streamwise locations, phases and FVD configurations.
 
\section{Conclusions and discussion} 
\label{sec:conclusions} 
The instability of supersonic boundary layers at Mach 5 over cooled walls and exposed to free-stream vortical disturbances has been studied by asymptotic and numerical methods. A biglobal temporal instability analysis has been performed and temporal-to-spatial transformations have been utilised to obtain the corresponding spatial growth rates. For typical flight conditions, nonlinear high-supersonic streaks induced by free-stream disturbances populate the boundary layer and stabilise the conventional non-radiating and radiating Mack modes. We have also identified secondary-instability modes, such as valley varicose modes as well as peak sinuous and varicose modes. Thanks to our unified receptivity--instability framework, we have established quantitative relations between the free-stream disturbance level and the characteristics of the streaks and the instability modes.

The valley varicose mode has been identified for the first time in a physically realisable streaky boundary layer. For the base-flow conditions and free-stream disturbance configuration examined here, this valley mode is the first secondary-instability mode to become unstable as the free-stream disturbance intensity is increased, appearing at a turbulence level $Tu \approx 0.03\%$, typical of flight and quiet wind-tunnel conditions. Unlike the previously reported peak modes, which concentrate in low-speed regions, it is concentrated in the high-speed valley regions, suggesting a possible role in the breakdown to turbulence of these regions. Furthermore, it occurs at low wall temperatures pertaining to thermal protection systems and causes intense temperature fluctuations near the wall. Its absence for \( T_w > 0.3T_{ad} \) may explain why it was not observed by \citet{paredes2019instability}, where the wall temperature was approximately $0.35T_{ad}.$ This mode could be of similar type to the instability mode pertaining to G\"ortler vortices in cooled-wall boundary layers \citep{brockhaus2024breakdown}. As shown in their figure 10, when the wall temperature decreases, transition moves upstream and the boundary layer is significantly distorted by disturbances near the wall, with a smaller streamwise wavenumber than that for an adiabatic wall. The valley mode has neither yet been measured experimentally, primarily because of limitations of measurement techniques at high Mach numbers very near the wall, nor has it been detected in DNS studies, where the receptivity of streaks to natural free-stream disturbances has not been taken into account. The inclusion of receptivity in DNS requires, in addition to a stringent spatial resolution, careful specifications of the free-stream flow in terms of (i) the far-field condition that faithfully characterises the interplay between the external disturbances and the boundary layer, and (ii) the inlet conditions that correctly represent the perturbation entrained into the boundary layer.

The high-Reynolds-number inviscid limit of the secondary-instability modes has also been examined. As reported in previous studies \citep*{li1995fundamental,xu_zhang_wu_2017,li2022secondary}, the peak sinuous and varicose modes do not approach the same inviscid limit, as they are primarily associated with the spanwise shear and the wall-normal shear, respectively. We have demonstrated that the new valley mode is symmetric and is therefore associated with the wall-normal shear, like the peak varicose mode. 

In our calculations, free-stream disturbances are assumed, for simplicity, to consist of a pair of oblique waves with a single frequency, and thus differ from those in a broadband disturbance environment. Under broadband forcing, the transition threshold may be lower than that obtained under the current assumption \citep{xu_zhang_wu_2017,xu2020gortler,xu2024excitation}. The present approach could be extended to include a full turbulence spectrum, as performed by \cite{zhang-etal-2011b} for incompressible flow. By combining the present results, which pertain to low-frequency disturbances, with those of \cite{dong-wu-2013} and \cite{wu-dong-2016}, which refer to order-one and higher-frequency disturbances, it would be possible to impose a more realistic broadband perturbation as inflow for DNS of transition in compressible boundary layers. However, such an approach would require reliable spectrum data of the incoming disturbance in the hypersonic regime that are not presently available due to substantial difficulties in the associated measurements (Duan \etal\ \citeyear{duan2019characterization}; Munoz \etal\ \citeyear{munoz2019freestream}), despite advances in experimental techniques \citep{gillespie2022multi,xiong2023characterization}. The broadband spectrum typical of realistic disturbances may also modify the amplitude thresholds for the onset of the secondary-instability modes and change the order of their appearance as the intensity of the free-stream disturbance grows. Nevertheless, the required amplitudes computed within our framework remain within physically plausible ranges for high-supersonic and hypersonic flight conditions in disturbed as well as relatively quiet atmospheric regions. Furthermore, our secondary-instability study is limited to the linear regime, which is important for understanding the initial stages of disturbance evolution. In more realistic scenarios, Mach wave emissions are also influenced by nonlinear interactions \citep{zhang2023generation,qin2024excitation,qin2025nonlinear} and turbulence \citep{duan2014numerical} in the boundary layer. The overall acoustic radiation from boundary layers undergoing nonlinear laminar-turbulent transition and from those in the fully-developed turbulent regime merits further study. Finally, a detailed analysis of fundamental, subharmonic and detuned modes, especially in the presence of broadband free-stream disturbance spectrum, is also of interest for future work.

\backsection{D.X. and P.R. wish to acknowledge the support of EPSRC (EP/T01167X/1). P.R. has also been supported by the AFOSR grant FA8655-21-1-7008  (International Program Office - Dr Douglas Smith).}

\backsection[Declaration of Interests]{The authors report no conflict of interest.}

\appendix
\refstepcounter{section} 
\label{app:check}
\section*{Appendix A. Fundamental and subharmonic modes}
\begin{figure}
\centering
\subfigure{
\includegraphics[width=0.42\textwidth]{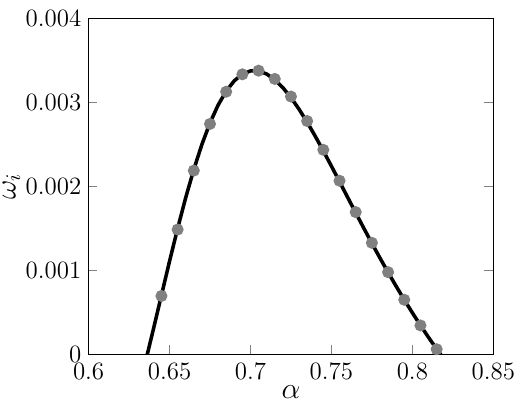}
\put(-165,118){($a$)}
\put(-71,4){\makebox(0,0){\colorbox{white}{\footnotesize $\alpha$}}}
\put(-160,70){\makebox(0,0){\rotatebox{90}{\colorbox{white}{\footnotesize $\omega_i$}}}}
}
\subfigure{
\includegraphics[width=0.42\textwidth]{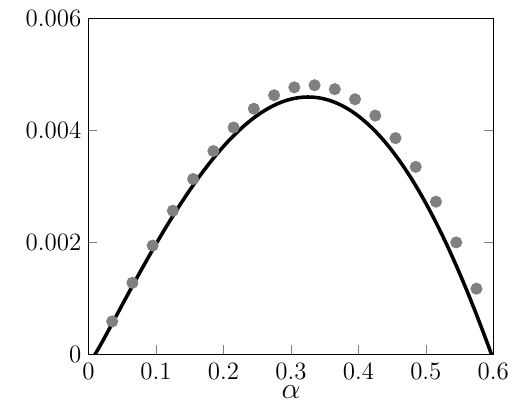}
\put(-165,118){($b$)}
\put(-71,4){\makebox(0,0){\colorbox{white}{\footnotesize $\alpha$}}}
\put(-160,70){\makebox(0,0){\rotatebox{90}{\colorbox{white}{\footnotesize $\omega_i$}}}}
}
\captionsetup{justification=raggedright}
\captionof{figure}{\color{black}Secondary instability of high-supersonic streaks: fundamental modes (solid line) and subharmonic modes (symbols). (a) Valley varicose modes at $r_t=2.5$ and ($b$) peak sinuous modes at $r_t=3.2$. The parameters are $\mathcal{M}_\infty=5,$ $\gamma=1.4$ and $Pr=0.707$.}
\label{fig:fon-sub}
\end{figure}

To complement the stability results in \S \ref{sec:results}, this appendix examines the sensitivity of the secondary-instability modes to the spanwise periodicity of the disturbances using a spanwise Floquet formulation. Figure \ref{fig:fon-sub} compares the temporal growth rates $\omega_i$ of fundamental and subharmonic disturbances for the valley varicose and peak sinuous modes studied in \S \ref{sec:second-family} as functions of streamwise wavenumber $\alpha$. Figure \ref{fig:fon-sub} shows that the growth rates are only weakly sensitive to the Floquet parameter. For this reason, only the fundamental modes are considered in the present study. A complete assessment of the modal dominance over the full range of Floquet parameters would require a more extensive analysis of the fundamental, subharmonic and detuned modes, which is beyond the scope of the present work.

\bibliographystyle{jfm}
\bibliography{dx}

\end{document}